\documentstyle[12pt]{article}

\def\hybrid{\topmargin -20pt  \oddsidemargin 0pt
      \headheight 0pt   \headsep 0pt
      \textwidth 6.25in 
      \textheight 9.5in 
      \marginparwidth .875in
      \parskip 5pt plus 1pt   \jot = 1.5ex}

\hybrid

\def\x{\times}
\def\ox{\otimes}
\def\o+{\oplus}
\def\ra{\rightarrow}
\def\lra{\longrightarrow}

\def\beqa{\begin{eqnarray}}
\def\eeqa{\end{eqnarray}}
                       
\newcommand{\un}{\underline}

\newcommand{\la}{\lambda}
\newcommand{\si}{\sigma}
\newcommand{\ga}{\gamma}

\newcommand{\C}{{\cal C}}
\newcommand{\D}{{\cal D}}
\newcommand{\E}{{\cal E}}
\newcommand{\F}{{\cal F}}
\newcommand{\G}{{\cal G}}
\newcommand{\cH}{{\cal H}}
\newcommand{\M}{{\cal M}}

\newcommand{\cP}{{\cal P}}
\newcommand{\cS}{{\cal S}}
\newcommand{\cO}{{\cal O}}

\newcommand{\resetcounter}{\setcounter{equation}{0}}

\begin{document}
\thispagestyle{empty}
\rightline{LMU-ASC 64/11}
\vspace{2truecm}
\centerline{\bf \LARGE Moduli restriction and Chiral Matter}
\vspace{.3truecm}
\centerline{\bf \LARGE in Heterotic String Compactifications}

\vspace{1.5truecm}
\centerline{Gottfried Curio\footnote{gottfried.curio@physik.uni-muenchen.de; 
supported by DFG grant CU 191/1-1}} 

\vspace{.6truecm}

\centerline{{\em Arnold-Sommerfeld-Center 
for Theoretical Physics}}
\centerline{{\em Department f\"ur Physik, 
Ludwig-Maximilians-Universit\"at M\"unchen}}
\centerline{{\em Theresienstr. 37, 80333 M\"unchen, Germany}}

\vspace{1.0truecm}

\begin{abstract}
Supersymmetric heterotic string models, built from 
a stable holomorphic vector bundle $V$ on a Calabi-Yau threefold $X$,
usually come with many vector bundle moduli whose stabilisation is a 
difficult and complex task.
It is therefore of interest to look for bundle constructions 
which, from the outset, have as few as possible bundle moduli.
One way to reach such a set-up is to start from a generic construction
and to make discrete modifications of it which are 
available only over a subset of the bundle moduli space.
Turning on such discrete 'twists' 
constrains the moduli to the corresponding subset of their moduli space:
the twisted bundle has less parametric freedom. 
We give an example of a set-up where this idea can be considered concretely. 
Such non-generic twists lead also to new contributions of chiral matter 
(which greatly enhances the flexibility in model building); 
their computation constitutes the main issue of this note.
\end{abstract}

\newpage

\section{Introduction}

A supersymmetric heterotic string model in four dimensions (4D) is given by 
the low energy effective theory arising from a compactification 
of the tendimensional heterotic string on a Calabi-Yau
threefold $X$ endowed with a polystable holomorphic vector bundle $V'$. 
Often one takes $V'=(V, V_{hid})$ with $V$ a {\em stable} bundle 
embedded in the visible $E_8$ whose commutant gives the unbroken gauge group in 4D
($V_{hid}$ plays the same role for the hidden $E_8$).
We restrict our attention to $V$ (and will assume $c_1(V)=0$).

Besides the Kahler and complex structure moduli of $X$ one gets 
moduli from the para\-meters of the bundle construction. As for the other 
moduli one searches for mechanisms, like world-sheet instantons and the
superpotential generated by them, to stabilise these moduli to particular 
values; at least one wants to restrict their freedom to certain subloci
of the moduli space, thereby simplifying the problem.
So it is of interest to have a bundle construction which, from the 
outset, comes with as few as possible bundle moduli.

One way to reach such a set-up is to start with a rather generic 
bundle construction and to make twists which are available only 
over a subset $\cS$ of the bundle moduli space $\M_V$: turning on such a twist
will restrict the moduli to $\cS$ if the twist is discrete.

We describe in the following a set-up where this idea can be 
considered concretely.
We emphasize from the outset that, although the moduli reducing effect
of the new twists is the rationale which lies behind our motivation to
consider them, we will focus in the present paper on another interesting
and phenomenologically relevant effect of the new twists.
Besides the issue of moduli stabilisation (or at least reduction)
the other prominent issue is the
influence of such twists on the cohomological invariants of $V$,
specifically the net generation number $N_{gen}=h^1(X,V)-h^1(X,V^*)=
-\chi(V)=-\int ch(V) Td(X)=-\frac{1}{2}c_3(V)$ (considered as number) 
and the impact for the 
anomaly cancellation condition involving $c_2(V)$ and $c_2(X)$. 
We will therefore compute in our concrete {\em set-up} the new contributions
to $c_3(V)$ and $c_2(V)$ provided by the mentioned twists
(and in the even more concrete {\em examples} of twists in this set-up 
which we will present we can evaluate the ensuing expressions even further).

Let us describe some related work. In the set-up we will choose, the case of spectral cover
bundles on elliptically fibered $X$, [\ref{FMW}] constitutes the basic reference. There
the more general possibility of using 'non-standard' twists in the sense described was already
seen (including, on a more implicit level, the corresponding issue of moduli restriction)
but, at the early stage of the investigations in the field at that time, the detailed 
discussion of the standard twist was sufficient for all purposes. 
Quite generally the point in this issue is to give worked out examples for non-generic twists 
(and making explicit the moduli reduction and the influence on the Chern classes) 
which was not in the focus of [\ref{FMW}]. The issue reappeared at the
surface on the occasion of investigations of dual $F$-theory models; in that language the 
question is discussed in [\ref{DW}] where also the general philosophy of using a non-standard
twist (with ensuing moduli reduction) is exemplified by a specific construction leading to a 
three-generation model. The examples given in the present paper are different, 
not just for the $SU(5)$ case and given directly in the heterotic set-up 
(though one can also use a 'heterotic language' directly in the $F$-theory set-up); furthermore 
computed is here not only the influence in the chiral matter expressed by the change in $c_3(V)$
but also the change in $c_2(V)$ (displaying 
also the specific parameter freedom in these Chern classes for the cases of our examples). 

It is interesting to note that the issue of moduli reduction by using such special
objects has not only be further explored in the $F$-theory context, for example in [\ref{cxstrfix2}]
(to quote just one paper from that direction of research); in a broader sense 
the issue converges also with another line of research in the heterotic context: 
in [\ref{cxstrfix}] heterotic constructions are made which exist only for a subset of the 
complex structure moduli, leading to a corresponding reduction of freedom in that moduli space.

{\em Structure of the paper}

To clarify the development of our argument let us point to a hierarchy of set-ups 
which become more and more concrete.
In {\em sect.~\ref{set-up}} we describe a concrete set-up 
which constitutes the first and most general layer; 
there we make the general idea of twisting concrete by 
pointing to the 'new' discrete twists which are possible
in the spectral cover scenario of bundle construction over an elliptically fibered
space $\pi: X\ra B$; here the mentioned twists can be handled effectively: 
we compute their impact on the Chern classes. 
Then in a second, already more concrete layer we specialise 
to certain general classes of 'new' (i.e.~non-generic) divisors on the spectral cover surface
and compute their new cohomological contributions, 
thus making our previous general formulae explicit for these cases
(whereas the moduli reduction effect in these examples is just suggested).
Our first example (second layer) is in {\em sect.~\ref{first example}}$\,$;
in {\em sect.~\ref{twist class}} we give another example 
of the type of twist class one can use in this set-up;
we also discuss the issue of moduli stabilisation (or rather restriction) in this connection.
In a third and final layer of concreteness we give 
in {\em sect.~\ref{Concrete examples n=4}} and {\em sect.~\ref{Concrete examples}}
explicit examples of the types of twist class described for the two most common
cases of $B$, the case of a Hirzebruch surface and of a 
del Pezzo surface, respectively,
thereby giving concrete examples of the general type of classes described in the second layer 
and evaluating our formulae for them. We conclude in sect.~\ref{Conclusions}.

\section{\label{set-up}A concrete set-up}

\resetcounter

Let us consider spectral $SU(n)$ vector bundles on an elliptic Calabi-Yau space
$\pi:X\ra B$ with section $\sigma$. (We will identify notationally 
$\sigma$, its image and the divisor and cohomology class of that image;
we also use the notation $c_1:=c_1(B)$, often with the pull-back to $X$
or $C$ understood; one has $\si^2=-c_1\si$, cf.~[\ref{FMW}].)

In this case one has 
\beqa
V=p_*(\cP\ox p_C^* L)
\eeqa
where one has the following objects
(this construction is by now fairly standard, cf.~[\ref{FMW}]):
one chooses a (ramified) $n$-fold cover surface $C\subset X$ over $B$, 
of cohomology
class $n\si + \pi^* \eta$ with\footnote{\label{base-point footnote}There are further conditions 
which have to be satisfied 
and have to be checked in detail in any concrete example (as we will do later).
Note first that the 
effectiveness of $C$ entails the effectiveness of $\eta$; furthermore the 
irreducibility of $C$ (which one needs to assume for the stability of $V$) is given just
for $\eta-nc_1$ effective and the linear system $|\eta|$ being base-point free;
the latter condition is best investigated further explicitly
on the different standard choices for the basis $B$: it holds on a Hirzebruch surface ${\bf F_k}$ 
if $\eta \cdot b\geq 0$ and on a del Pezzo surface ${\bf dP_k}$ 
with $2\leq k \leq 7$ if $\eta \cdot E\geq 0$ for all curves $E$ with 
$E^2=-1$ and $E\cdot c_1=1$ (such curves generate the effective cone) 
(for the notation used cf.~sect.~\ref{Concrete examples} where this information is used;
similarly also in sect.~\ref{Concrete examples n=4}). 
We remark further that one has also $c_2(X)=12c_1\si+11c_1^2+c_2(B)$ (cf.~[\ref{FMW}]).}
$\eta\in H^{1,1}(B)$, and a line bundle
$L$ over $C$; $\cP$ is the Poincare bundle over $X_{(1)}\x_B X_{(2)}$ 
restricted here to $X\x_B C$ and $p$ and $p_C$
the projections to the first and second factor, respectively
(here one has
$c_1(\cP)=\Delta-\si_1-\si_2-c_1$ with the diagonal class $\Delta$ in the
fibre product and the corresponding section classes from the factors; 
all necessary pull-backs are understood).

The condition $c_1(V)=0$ will fix $c_1(L)$ in $H^{1,1}(C)\cap H^2(C, {\bf Z})$
up to a class $\ga$ in $ker (\pi_{C*})$:
\beqa
c_1(L) &=& \frac{n\si + \eta + c_1}{2}+\ga
\eeqa
where one has $\pi_{C*}\ga=0$
(here $\pi_C: C \ra B$ is the restricted projection; we will usually suppress the pull-back notation
and write just $\phi$ for $\pi^* \phi$ or $\pi_C^* \phi$).

The equation for $C$ is given by 
\beqa
w&=&a_0 z + a_2 x + a_3 y=0\\
w&=&a_0 z^2 +a_2 xz + a_3 yz + a_4 x^2 +a_5 xy=0
\eeqa 
for $n=3$ and $n=4$ or $5$, resp.~(with $a_5=0$ for $n=4$; 
here $x,y,z$ are Weierstrass coordinates of the elliptic fibre
and $a_i$ sections of suitable line bundles over $B$).

\newpage

\noindent
If one assumes that $C$ is ample one has $H^{1,0}(C)=0$ and $L$ is 
determined by its first Chern class (no further continuous moduli occur);
then also the curve $A_B:=C\cap B\subset B$ is ample
($A_B$, or $A_C$ later, will also denote the cohomology class). We will,
however, have reason to consider also the case that this curve, and thus
$C$ as well, is not ample (cf.~sect.~\ref{Concrete examples}).
In this case further, continuous degrees of freedom, related to 
$H^1(C, \cO)/H^1(C, {\bf Z})$, occur which are fibered over the discrete
classification of the line bundles provided by the Chern class.
We nevertheless continue to speak of {\em discrete} twists also then,
as the important point for us is that (regardless of the additional continuous
degree of freedom in the fibre of this situation) the use of a 'new' twist
(not belonging to the standard twists available generically) can not be turned
off continuously, i.e.~the effect of reduction in the vector bundle moduli 
space, in which we are interested, takes place in any case.

\subsection{The standard situation}

If one wants to describe the possible freedom 
one has in choosing $\ga$, one can say generically
only the following: the only obvious classes on $C$ are, besides the
section $\si|_C$, the pull-back classes $\pi^* \phi$ where the class $\phi$ comes
from the base. One finds [\ref{FMW}] that 
$\pi_{C*} \si|_C = \bar{\eta}:=\eta - n c_1$ and so the only class in
$ker (\pi_{C*})$ available in general is
\beqa
\label{standard ansatz}
\ga &=& n\si|_C - \pi_C^* \bar{\eta}
\eeqa
(or suitable multiples $\la \ga$ of it; at this point 
an integrality issue occurs\footnote{$\la$ has to be half-integral 
in a specific way depending on the parity of $n$ ($n$ odd needs $\la\in \frac{1}{2}+{\bf Z}$
while $n$ even needs $\eta\equiv c_1 \, (2)$ for $\la \in {\bf Z}$ and $0\equiv c_1 \, (2)$
for $\la \in \frac{1}{2}+{\bf Z}$)} 
which we do not need to make explicit here; important is that $\la$ has only discrete freedom). 

One gets the following formulae (cf.~[\ref{FMW}] and [\ref{C}]; for $p: X\x_B C \ra X$ cf.~above)
\beqa
c_2(V) &=& \eta \si - \frac{n^3-n}{24}c_1^2-\frac{n}{8} \eta \bar{\eta}
- \frac{1}{2}\pi_{C*} \ga^2 \\
\frac{1}{2}c_3(V) &=& \frac{1}{2}p_*\Big(\ga c_1^2(\cP)\Big)
\eeqa
 
In this form the formulae hold for a {\em general} $\ga$. 
With the {\em concrete generic}
$\ga$ class given above one finds\footnote{this corresponds to the choice 
$\la=1$; for $n$ odd $\la$ has to be strictly
halfintegral, but it is obvious how the expressions have to be 
adapted: (\ref{standard contribution 1}) and (\ref{standard contribution 2}) 
come with a factor $\la^2$ and $\la$, resp., in general}$^{,}$\footnote{the final term 
on the right hand side of (\ref{standard contribution 2})
actually occurs at first as $\si \eta \bar{\eta}$, 
i.e.~$\si \pi^*\eta \pi^*\bar{\eta}$; 
as this is interpreted in any case as a number
one can simply read it as intersection number on $B$}
\beqa
\label{standard contribution 1}
\pi_{C*}\ga^2&=&-n \eta \bar{\eta} \\
\label{standard contribution 2}
\frac{1}{2}p_*\Big(\ga c_1^2(\cP)\Big)&=&\eta \bar{\eta}
\eeqa

\subsection{The new, extended class of twists}

Now let us assume that, at least for a certain subset ${\cal S}$ of the moduli
space $\M_V$, further divisor classes on $C$ exist
(such that further corresponding cohomology classes, 
denoted by $\tilde{\chi}$ below, in the expression for $\ga$ can occur). 
Then we can make a more general ansatz for the cohomology class $\ga$ 
(where $\rho$ here is still a class coming from the base)\footnote{to avoid unclear notation 
we now denote the pull-back class by $\rho$ instead of the former $-\bar{\eta}$ 
(cf.~(\ref{standard ansatz})); the pull-back operation itself is suppressed, so $\rho$ is actually 
$\pi_C^* \rho$; if no confusion can arise we will also 
suppress in the following the restriction and write just $\si$ for the class $\si|_C$} 
\beqa
\label{preliminary gamma ansatz}
\ga &=& n\si + \rho +\tilde{\chi}
\eeqa
The condition $\pi_{C*}\ga=0$ amounts now to 
$n(\bar{\eta}+\rho)+\pi_{C*}\tilde{\chi}=0$; 
to secure the divisibility of $\pi_{C*}\tilde{\chi}$ by $n$ 
we are led to the slightly modified ansatz $\tilde{\chi}:=n\chi$, 
that is
\beqa
\label{gamma ansatz}
\ga &=& n (\chi + \si) + \rho = n(\chi + \si)- \pi_{C*}(\chi+\si)
\eeqa
In the last rewriting we made manifest the condition on $\rho$
which guarantees\footnote{To avoid any confusion 
note that the final term $\pi_{C*}(\chi+\si)$ is, in itself, a class projected down to $B$; 
if it occurs, as it is the case here, in a formula for a class on $C$ (the class $\ga$),
then this means that it has to be read as being pulled-back to $C$; in
other words this means actually the class $Q:=\pi_C^*\pi_{C*}(\chi+\si)$
(for $\chi=0$ one gets back $Q=\pi_C^* \bar{\eta}$, 
cf.~(\ref{standard ansatz}));
so both terms in the final expression $P-Q$ on the right hand side of 
(\ref{gamma ansatz}) fulfil $\pi_{C*}P=n\pi_{C*}(\chi+\si)=\pi_{C*}Q$, 
thus giving indeed $\ga \in ker (\pi_{C*})$.}
$\ga \in ker (\pi_{C*})$
(in the final expression one can also turn off, discretely, $\chi$ to get
back (\ref{standard ansatz})).
Again one may also consider suitable multiples $\la \ga$ and
an integrality issue occurs\footnote{\label{integrality footnote}$\la$ has to be half-integral 
in a specific way depending on the parity of $n$ 
($n$ odd needs $\la\in \frac{1}{2}+{\bf Z}$ and $\chi\equiv 0 \, (2)$ 
while $n$ even needs $\eta\equiv c_1 \, (2)$ for $\la \in {\bf Z}$ 
and $\pi_{C*}\chi \equiv c_1 \, (2)$ for $\la \in \frac{1}{2}+{\bf Z}$)}.

Now we are interested in the new contributions to the Chern classes arising
from the {\em new} class $\chi$, i.e., from the class which is {\em not} 
already contained in the span of the classes which are generically present 
(which consist, besides the 
special class $\si$ (i.e.~$\si|_C$), in the pull-back classes $\pi^* \phi$).
It is useful to recall in this context the 'projection formula' 
$\pi_{C*}(\pi_C^* \phi  \cdot \si)=\phi \cdot \pi_{C*}\si$
involving pull-back classes.
As one has $\pi_{C*} \pi_C^* \phi = n \phi$ 
one can write then also
$n \; \pi_{C*} (\pi_C^* \phi \cdot \si)=\pi_{C*} \pi_C^*\phi \cdot \pi_{C*}\si$.
Therefore, from classes $\chi$
(which are {\em not} pull-back classes like $\pi_C^* \phi$)
one can expect, as new contributions,
non-zero terms built from a corresponding difference of the
right and the left hand side of this relation, i.e. terms like 
$\pi_{C*} \chi \cdot \pi_{C*}\si - n \; \pi_{C*}( \chi \cdot \si)$,
or, more generally, 
$\pi_{C*} \chi \cdot \pi_{C*}\zeta - n \; \pi_{C*}( \chi \cdot \zeta)$
(where the further class $\zeta$ on $C$ could be $\chi$ itself, for example,
cf.~(\ref{c2 formula}) below).

One gets now indeed
\beqa
\pi_{C*}\ga^2 &=& 
-n \Bigg[ \Big(\pi_{C*}(\chi + \si)\Big)^2-n \;\pi_{C*} (\chi + \si)^2 \Bigg]\\
\label{c2 formula}
&=& -n \Bigg[ \pi_{C*} \si \cdot \pi_{C*} \si  - n\; \pi_{C*} \si^2 \nonumber\\
&&\;\;\;\;\;\;\; 
+2\Big( \pi_{C*}\chi \cdot \pi_{C*} \si  - n\; \pi_{C*} (\chi \cdot \si)\Big)
\nonumber\\
&&\;\;\;\;\;\;\;  + \pi_{C*}\chi \cdot \pi_{C*}\chi - n\; \pi_{C*}\chi^2 \Bigg]
\eeqa
Note that here the first line in the big brackets on the right hand side in 
(\ref{c2 formula}) is the usual term 
$\bar{\eta}\bar{\eta}+nc_1 \bar{\eta} = \eta \bar{\eta}$, 
cf.~(\ref{standard contribution 1}). 
The additional, new contributions in the last two lines are
now indeed of the expected form for which we argued in the previous paragraph.

And similarly one gets (using $\si_i \cdot c_1(\cP)=0$)
\beqa
c_3(V)&=&p_*\Bigg(\ga c_1^2(\cP)\Bigg)=
p_*\Bigg((\rho + n \chi)\Big(-2\si_2\si_1+c_1(-3\Delta+\si_1+\si_2)\Big)\Bigg)\\
&=&\Big( -2\rho(\bar{\eta}+nc_1)\, \si_1
-2n\pi_{C*}(\chi\si_2 + \chi c_1)\; \si_1\Big)
\eeqa
This leads, after using $\rho=-\bar{\eta}-\pi_{C*}\chi$, 
to the further evaluation
\beqa
-N_{gen}&=&-\rho \eta -n\pi_{C*}(\chi\, (\si+c_1))\\
&=&\eta\bar{\eta}+\eta \, \pi_{C*}\chi -n\; \pi_{C*}(\chi\, (\si+c_1))\\
&=&\eta\bar{\eta}+\bar{\eta}\, \pi_{C*}\chi -n\; \pi_{C*}(\chi \si)
\eeqa
This gives the final formula for the generation number which shows that 
the new contribution is just of the structurally expected type
(cf.~(\ref{standard contribution 2}))
\beqa
\label{Ngen formula}
-N_{gen}&=&
\eta\bar{\eta}+\pi_{C*}\chi\cdot \pi_{C*}\si - n\; \pi_{C*}(\chi\cdot \si)
\eeqa

So let us finally list (using again $\pi_{C*}\si=\bar{\eta}=\eta-nc_1$)
the complete expressions one gets if one turns on,
as specified in (\ref{gamma ansatz}), 
a non-pull-back class $\tilde{\chi}=n\chi$ in the twist
\beqa
\label{c2 contribution}
c_2(V)&=&\eta \si -\frac{n^3-n}{24}c_1^2-\frac{n}{8}\eta\bar{\eta}\nonumber\\
&&
+n\la^2\Bigg[ \frac{1}{2}\eta\bar{\eta}+\pi_{C*}\chi\cdot \pi_{C*}\si - n\; \pi_{C*}(\chi\cdot \si)
+\frac{1}{2}\Big(\pi_{C*}\chi \cdot \pi_{C*}\chi - n\; \pi_{C*}\chi^2\Big)\Bigg]\;\;\;\;\;\;\;\;\;
\\
\label{Ngen contribution}
-N_{gen}&=&
\la\Big[ \eta \bar{\eta} + \pi_{C*}\chi\cdot \pi_{C*}\si - n\; \pi_{C*}(\chi\cdot \si)\Big]
\eeqa
(the first terms in the $[ ... ]$ brackets are the standard terms, the rest the corrections).

In the remaining sections we want to give examples of this construction, i.e.~concrete
classes to twist with and the corresponding evaluation of the new cohomological
contributions; furthermore we want to make remarks on the issue of moduli reduction. 
But before we come to this let us consider two related issues:
the direct chiral matter computation of $N_{gen}$ and 
the set of classes which are available in general for $\chi$.

\subsection{The direct chiral matter computation of $N_{gen}$}

Let us first recall (cf.~[\ref{C}]) the computation of $N_{gen}$ 
from the net amount $h^1(X, V)-h^1(X, V^*)$ of chiral matter 
for the standard $\ga$ twist. 
This proceeds, as $H^1(X, V)$ is localised along $\pi^*A_B$ 
and by noting that $V|_B\cong \pi_{C*}L$, 
with the help of the Leray spectral sequence 
(which itself simplifies because of $R^0\pi_* V=0$)
\beqa
0\lra H^1(B, R^0\pi_* V) \lra H^1(X, V) \lra H^0(B, R^1\pi_*V)
\ra H^2(B, R^0\pi_* V)
\eeqa
One computes (taking into account the relative Serre duality 
$(R^1\pi_*V)^* \cong \pi_* (V^*\ox K_B^*)$)
\beqa
H^1(X, V)&\cong & H^0(B, R^1\pi_* V)\; \cong \;
H^0(A_B, R^1\pi_* V|_{A_B})\\
&\cong & H^0\Big(A_B, \Big[ L|_{A_C}\ox \pi_C^* K_B|_{A_C}\Big]_{A_B}\Big)
\eeqa 
where the internal brackets in the final expression indicate 
that the line bundle inside them, which a priori lives on $A_C:=\si|_C$,
is interpreted on $A_B$.
One gets for $N_{gen}$ the result (we put again $\la =1$; the brackets 
with subscript $B$ indicate that the intersection product on $C$ inside them
is interpreted afterwards as an intersection product on $B$)
\beqa
\chi\Big(A_B, \Big[ L|_{A_C}\ox \pi_C^* K_B|_{A_C}\Big]_{A_B}\Big)\!\!\!\!\!\!
&=& \!\!\!\! -\frac{1}{2}\deg K_{A_B} + \deg [ L|_{A_C}]_{A_B}+\deg K_B|_{A_B} 
= \Big[ \ga \cdot A_C \Big]_B\;\;\;\;\;\;\;\;\; \\
&=& \!\!\!\!  \Big[ \ga \cdot \si|_C \Big]_B = \Big[ -\eta \cdot \si|_C \Big]_B  
= -\pi_{C*}(\eta \cdot \si|_C)\\
&=& \!\!\!\!   -\eta\bar{\eta}
\eeqa
where we have inserted the relation 
(we have also used $\deg K_A=\deg K_C|_A+\deg K_B|_A$)
\beqa
\deg [L|_{A_C}]_{A_B}=
\deg L|_{A_C}&=&
\frac{1}{2}\Big(\deg K_C-\deg K_B\Big)\Big|_{A_C}+ \ga \cdot A_C \\
&=&\frac{1}{2}\deg K_A -\deg K_B|_A +\ga \cdot A_C
\eeqa
(if the curve $A$ does not carry a subscript, indicating in which surface,
$C$ or $B$, is has to be interpreted, then that does not matter).

Now in the new, more general case one gets
\beqa
\ga \cdot \si&=&
n\chi \cdot \si - nc_1 \si -(\pi_C^*\pi_{C*}\chi)\si - \bar{\eta}\si\nonumber\\
&=&n\chi \cdot \si -(\pi_C^*\pi_{C*}\chi)\si -\eta \si
\eeqa
The latter expression projects under $\pi_{C*}$ down to
\beqa
\pi_{C*}(\ga \cdot \si)&=&n\pi_{C*}(\chi \cdot \si) 
- \pi_{C*}\chi\cdot \pi_{C*}\si -\eta\bar{\eta}
\eeqa
Thus, for $-N_{gen}$, we arrive again at the expression (\ref{Ngen formula}).

\subsection{Remarks on the classes available for $\chi$}

Let us quantify the available resources for classes like $\chi$.
From the outset one has just the class $\si|_C$ and the pull-back classes
$\pi_C^* \phi$ at one's disposal; so 
the number of classes which are available generically is
\beqa
1+h^{1,1}(B)&=& e(B)-1
\eeqa
(if one makes furthermore use of the fact that $B$ is
a rational surface of $1=p_g(B)=1-h^{1,0}(B)+h^{2,0}(B)=(c_1^2+e(B))/12$,
using Noether's formula,
one obtains here the alternative evaluation $11-c_1^2$).

On the other hand the number of classes available in principle is
given by the rank of the lattice $H^2(C, {\bf Z})\cap H^{1,1}(C)$; here
one computes\footnote{using  $c_1(C)=-(n\si+\eta)|_C$ and $c_2(C)=C^2|_C+c_2(X)|_C$, 
cf.~[\ref{FMW}], which give $c_1^2(C)=3n\eta\bar{\eta}+n^3c_1^2$ and 
$e(C)=3n\eta\bar{\eta}+(n^3-n)c_1^2+12\eta c_1+ne(B)$, from which one derives
in turn, using Noether's formula now applied to $C$, that
$h^{2,0}(C)-h^{1,0}(C)=
\frac{n}{2}\eta\bar{\eta}+\frac{n^3-n/2}{6}c_1^2+\eta c_1+\frac{n}{12}e(B)-1$}
for $H^{1,1}(C)$ itself
\beqa
h^{1,1}(C)
&=&2n\eta\bar{\eta}+\frac{4n^3-5n}{6}c_1^2+10\eta c_1+\frac{5n}{6}e(B) + 2 h^{1,0}(C)
\eeqa
(if one makes again use of the fact that $B$ is a rational surface 
one obtains here the alternative evaluation
$2n\eta\bar{\eta}+\frac{4n^3-10n}{6}c_1^2+10\eta c_1+10n + 2 h^{1,0}(C)$).
It now depends on the complex structure of $C$ which of these forms
have integral periods when integrated against a basis of integral cycles,
and thus belong also to $H^2(C, {\bf Z})$. In our situation the actually
available complex structures of $C$ come from its 'motions' in the ambient 
space $X$, i.e.~from the possible different equations (up to an overall rescaling)
for $C$ in $X$; this, as described, comprises just the continuous
part of the moduli space $\M_V$ of the bundle.

One can view the problem to determine the intersection 
$H^2(C, {\bf Z})\cap H^{1,1}(C)$ also from the other side:
the demand that a topological class $\xi\in H^2(C, {\bf Z})$ has type $(1,1)$
(so is related to a holomorphic cycle) is expressed by the orthogonality
$\xi \perp \alpha$ for all $\alpha$ in a base of the subspace which 
constitutes $H^{2,0}(C)$. A priori each $\alpha$ could be everywhere in\\
\noindent
$U\! =\! \{\beta\! \in \! H^2(C, {\bf R})|
\beta \wedge \beta \! =\! 0, \beta \wedge \bar{\beta}\! >\! 0\}$. 
So classes $\! \xi\! $ are relatively scarce (cf.~[\ref{FMW}], sect.~7.4).\\
\indent
A well-known analogous situation is that of a $K3$ surface 
where the demand for a higher rank (the Picard number) of the span 
of the sought-after classes $\xi$ restricts one accordingly in the
moduli space.\footnote{But note that in that example the relevant moduli space
are the possible positions of $H^{2,0}(K3)$ in ${\bf P}(U)$; by contrast
in our case those moduli of $C$ 
which are relevant for the spectral bundle set-up 
are not exactly the internal complex structures of $C$ but the external 
motions in $X$, i.e. $H^{2,0}(C)$ itself.}

\section{\label{first example}A first example of a  non-generic twist class}

\resetcounter

In the two examples of a non-generic twist class 
given in the present and in the next chapter
we will use the idea that under special conditions on the (bundle) moduli
one of the generically present classes $\pi_C^* \phi$ and $\si|_C$ becomes reducible; 
then a component of this reducible class represents a 'new' class to twist with.

In this section we take the first case: we will look for a case where under certain conditions
the preimage $\C:=\pi_C^{-1}(c)$ of a curve $c\subset B$ becomes reducible in $C$
\beqa
\label{vertical decomposition}
\C&=&\C_1+\C_2
\eeqa

We will assume $n>3$ and to be as concrete as possible
we choose the cases $n=4$ or $n=5$ (which are also phenomenologically the most important ones;
the factorisation idea described in the following can be analogously considered for $n>5$).
The spectral cover equation is 
(with $a_5=0$ for $n=4$; here $a_i$ are global sections 
of\footnote{with a common abuse of notation to denote divisors
by symbols for corresponding cohomology classes} 
$\cO_B(\eta-ic_1)$)
\beqa
w&=&a_0 z^2 +a_2 xz+a_3 yz + a_4 x^2 +a_5 xy
\eeqa

Now let us consider in a first, {\em preliminary} step the following factorisation
\beqa
\label{global factorization}
w&=&(f_1 z + g_1 x + h_1 y) \, (f_2 z + g_2 x) 
\eeqa
(with $h_1=0$ for $n=4$)
where $f_1, g_1, h_1, f_2, g_2$ are sections of suitable line bundles over $B$:
if one denotes the vanishing divisor of $g_2$, say, by $(g_2)$ one has
\beqa
f_1&\in & H^0\Big(B, \cO_B(\eta-2c_1-(g_2))\Big)\\
g_1&\in & H^0\Big(B, \cO_B(\eta-4c_1-(g_2))\Big)\\
h_1&\in & H^0\Big(B, \cO_B(\eta-5c_1-(g_2))\Big)\\
f_2&\in & H^0\Big(B, \cO_B(2c_1+(g_2))\Big)\\
g_2&\in & H^0\Big(B, \cO_B((g_2))\Big)
\eeqa
The relations to the original coefficients are
\beqa
a_0&=&f_1 \, f_2\\
a_2&=&f_1 \, g_2 + g_1 \, f_2\\
a_3&=&h_1\, f_2\\
a_4&=&g_1\, g_2\\
a_5&=&h_1\, g_2
\eeqa
If the original coefficients $a_i$ can be written in this rather special way
one gets the relation 
\beqa
\label{factorization relation}
a_0a_5^2-a_2a_3a_5+a_3^2a_4&=&0
\eeqa
($a_3\! =\! 0$ for $n\! =\! 4$).$\!$
Note that this means here identical vanishing over all of $B$. If considered as an equation for
a curve in $B$ it describes [\ref{FMW}] the localization curve of the bundle $\Lambda^2 V$.

We have not yet considered the question whether the relation (\ref{factorization relation}),
which as we showed is necessary to have a factorization like (\ref{global factorization}),
is also sufficient to have such a relation. 
We will consider the question further in a moment
in the somewhat reduced framework of factorization 
in which we are actually interested and to which we turn now.

The factorizability considered above is much more then we actually have to demand.
Let $c$ denote a (smooth irreducible reduced) curve in $B$ and assume the following
factorisability of $w$ over the elliptic surface $\E_c := \pi^{-1}(c)$
(with $F_1  := f_1|_c$ and so on)
\beqa
\label{factorization over c}
w|_{\E_c}&=&(F_1 z + G_1 x + H_1 y) \, (F_2 z + G_2 x) \, |_{\E_c}
\eeqa
where $F_1, G_1, H_1, F_2, G_2$ are now sections of suitable line bundles
over $c$: for example one has that $F_1\in H^0\Big(c, \cO_c((\eta-2c_1)|_c-(G_2))\Big)$
and $A_0=F_1\, F_2$ (with $A_i:=a_i|_c$) and analogous expressions for all the other equations.

So one gets now as condition for the factorizability over $c$ that the 
curve given in\footnote{which is now read as an equation
for a curve in $B$ and not as an identical vanishing over all of $B$} 
(\ref{factorization relation}) has $c$ as a component, 
i.e.~that the equation (\ref{factorization relation}) is fulfilled along $c$
(the concrete case in which we are interested is $c\cong {\bf P^1}$ 
which we assume now for simplicity)
\beqa
\label{factorization relation over c}
A_0A_5^2-A_2A_3A_5+A_3^2A_4&=&0
\eeqa
Note that the relation just presented is not only necessary 
but also sufficient to have (\ref{factorization over c}) (we assume here $n=5$).
Note first that because of (\ref{factorization relation over c}) one has $A_5|A_3A_4$,
so that one can write $A_5=H_1 G_2$ with $H_1|A_3$ and $G_2|A_4$; 
let us write furthermore $A_4=G_1G_2$ and $A_3=H_1F_2$.
From (\ref{factorization relation over c}) one gets $A_5F_2=A_3G_2|A_0A_5$, 
such that $F_2|A_0$ and one can write $A_0=F_1F_2$. 
From these determinations it follows already,
once more with (\ref{factorization relation over c}), that 
$A_2=(A_0A_5^2+A_3^2A_4)/A_3A_5=F_1G_2 + G_1F_2$.

If condition (\ref{factorization over c}) is fulfilled 
one has the decomposition (\ref{vertical decomposition})
with $\C_1$ and $\C_2$ corresponding to the first and second factors 
in (\ref{factorization over c}), respectively: in other words the five-fold cover $\C$ of $c$
decomposes into a triple cover $\C_1$ and a double cover $\C_2$
(this is for $n=5$; for $n=4$ one has to adjust these assertions, cf.~sect.~\ref{case n=4}).

For future reference we note the relation\footnote{where the divisor $A_B$ and the zero divisors 
$(a_5), (h_1)$ and $(g_2)$ denote also the cohomology classes}
(until sect.~\ref{case n=4} we assume now $n=5$)
\beqa
\label{trivial relation}
A_B=\bar{\eta}=(a_5)=(h_1)+(g_2)
\eeqa

To compute the contributions in (\ref{c2 contribution}) and (\ref{Ngen contribution}) 
for $\chi=\C_2$, say, it remains, having $\pi_{C*}\C_2=2c$, 
to compute $\pi_{C*}(\C_2 \cdot \si|_C)$ and $\pi_{C*}(\C_2^2)$.
For this note first that $\C_1\cdot \si|_C+\C_2\cdot \si|_C=\C\cdot \si|_C=\E_c|_C \cdot \si|_C=
\E_c\cdot \si\cdot C=\E_c|_{\si}\cdot C|_{\si}=c\cdot A_B
=(h_1)\cdot c+(g_2)\cdot c =\deg H_1 + \deg G_2$
because of (\ref{trivial relation}); in particular one has $\C_2\cdot \si|_C=(g_2)\cdot c$.

To compute $\pi_{C*}(\C_2^2)$ let us compute first $\C_1\cdot \C_2$
(intersection number in $C$).
For this note the following determinations of the cohomology classes of involved 
divisors\footnote{In (\ref{C_2 equation}) the equation in each fibre plane 
${\bf P^2_{x,y,z}}$ is linear in the Weierstrass coordinates, 
so intersects the elliptic cubic {\em three} times; 
only two of these fibre points carry information (the fibre points $q_1, q_2$ of $\C_2$), 
a third one lies always at the zero point $p_0$:
for $f_2 z + g_2 x = z (f_2 + g_2 \frac{x}{z})$ shows as divisor three zeroes at $p_0$ from $z$
and two zeroes at $q_1, q_2$ and a double pole at $p_0$ from the affine part;
by contrast for $\C_1$ a {\em triple} pole cancels the zeroes of $z$ 
while the affine part has three relevant zeroes (the fibre points of $\C_1$).}
\beqa
\label{C_1 equation}
\C_1              \; = \; (f_1 z + g_1 x + h_1 y)|_{\E_c}&=&
\Big(3\si+ \eta- 2c_1 - (g_2)\Big)\Big|_{\E_c}\\
\label{C_2 equation}
\C_2 + \si|_{\E_c}\; = \; (f_2 z + g_2 x)|_{\E_c}        &=&
\Big(3\si      + 2c_1 + (g_2)\Big)\Big|_{\E_c}
\eeqa
(divisors in the surface $\E_c$).
Thus one gets (as intersection number in $\E_c$ and also\footnote{assuming 
that $\C_1$ and $\C_2$ have no common component 
such that no self-intersection number is involved} in $C$)
\beqa
\label{intersection evaluation}
\C_1\cdot \C_2 &=& \Big( 2 \bar{\eta}+6c_1+(g_2)\Big) c
\eeqa
Additional standard assumptions which are often adopted (though not strictly necessary) 
are that $\bar{\eta}$, which is effective, is even ample.
If one assumes furthermore $c_1$ effective one excludes from the 
standard examples\footnote{Hirzebruch surfaces ${\bf F_k}$ ($k=0,1,2$), del Pezzo surfaces
${\bf dP_k}$ ($k=0, \dots ,8$) and the Enriques surface}
for $B$ only the Enriques surface;
and if one assumes $c_1$ even to be ample one excludes only ${\bf F_2}$ in addition.
Making these assumptions the first two terms on the right hand side of 
(\ref{intersection evaluation}) are $>0$; so for $\C_1\cdot \C_2=0$ 
one then would need $(g_2)\, c<0$, 
in particular neither of the effective divisors $(g_2)$ and $c$ could be ample.

With this information and the projection formula
$\pi_{C*}(\C_2\cdot \pi_C^{*}\, c)=\pi_{C*}(\C_2)\cdot c$ we get finally 
(note that for $\pi_{C*}(\C_1^2)$ one gets $3c$ instead of $2c$ as first term in the final bracket)
\beqa
\pi_{C*}(\C_2^2)
&  = &  \pi_{C*}(\C_2\cdot \C - \C_2\cdot \C_1)
\; = \; \Big( 2c - 2 \bar{\eta} - 6c_1 - (g_2)\Big) c
\eeqa
Thus one gets finally from (\ref{c2 contribution}), (\ref{Ngen contribution}) 
in this example of $\chi=\C_2$ (as cohomology class) the complete expressions
(the first terms in the $[\, ...\, ]$ brackets on the right hand sides are the standard 
contributions, the terms proportional to $c$ are the new contributions) 
\beqa
\label{first new c_2}
c_2(V)\!\!\!&=&\!\!\!\eta \si - 5c_1^2-\frac{5}{8} \eta \bar{\eta}+
5\la^2  \Bigg[\frac{1}{2}\eta\bar{\eta} 
+ \Bigg(2\bar{\eta}-5(g_2)-3c+5\Big( \bar{\eta}+3c_1+\frac{1}{2}(g_2)\Big)\Bigg)c\Bigg]
\;\;\;\;\;\;\;\;\;\\
\label{first new Ngen}
-N_{gen}\!\!\!&=&\!\!\!\la \Big[\eta \bar{\eta}+ \Big(2\bar{\eta}-5(g_2)\Big)c\Big]
\eeqa

Taking $\chi= \C_1$ instead of $\C_2$ gives, 
with $(3\bar{\eta}-5(h_1))c$ as new term in (\ref{first new Ngen}), 
by (\ref{trivial relation}) the negative of the present new term
(the same term $(3\bar{\eta}-5(h_1))c$ just replaces for $\chi=\C_1$ 
the term $(2\bar{\eta}-5(g_2))c$ for $\chi=\C_2$
in (\ref{first new c_2})).

One also has to take into account the parity considerations 
(cf.~footn.~\ref{integrality footnote}). 
Here, in our case of $n=5$, 
one has to check whether $\chi=\C_1$ or $\C_2$ is even when considered in the surface $C$.
Now note first that the curve $\C$, in whose components $\C_1$ and $\C_2$
we are interested, can be considered as a curve either in the spectral cover surface $C$
or in the elliptic surface $\E_c=\pi^{-1}(c)$: the representation as a divisor in these cases
reads $\C=\E_c|_C$ and $\C=C|_{\E_c}$, respectively.
What one finds immediately from (\ref{C_1 equation}) and (\ref{C_2 equation}) is
that {\em considered on the surface $\E_c$}
only the class $\chi=\C_2$ can be seen to be even and actually is so for $\deg G_2$ even.
This is, however, not related directly to the issue of being even on $C$.
A {\em necessary} condition at least for the latter fact is that the curve class in question
is even when {\em considered in the threefold $X$}.
Here one finds from 
$\C_1=(3\si + \eta - 2c_1-(g_2))\cdot \pi^{-1}(c)$ and $\C_2=(2\si +2c_1+(g_2))\cdot \pi^{-1}(c)$
of curve classes in $X$ (where $\pi$ is the projection from $X$ to $B$ 
whereas $\pi_C$ is the projection from $C$ to $B$) 
that this certainly holds if the class $c$ in $B$ is even or, in the case of $\C_2$, if 
$(g_2)\cdot c=\deg G_2$ is even. 
But none of these conditions gives a sufficient condition for evenness on $C$ 
of the curve class in question. 
The situation will be better in the case of $n=4$ considered below in sect.~\ref{case n=4}.

Another issue is whether one has to demand 
that the components $\C_1$ and $\C_2$ of $\C$ do not intersect 
to make sure the smoothness of $C$ (this is just to be on the save side; 
the spectral cover construction may make sense also in more general cases).
The decomposition 
\beqa
C\cap \E_c= \C_1+\C_2
\eeqa
leads one to expect the picture that $C$ 
decomposes {\em near $\E_c$} in two {\em local} branches given by a triple and a double cover
(globally $C$ will of course generically be irreducible). Potential intersection points
of the two local branches do not necessarily have to be interpreted as a curve of double points 
of $C$ as one does expect in any case ramification points of the covering 
$\pi_C :  C\ra  B$.
Despite the fact that double points are also possible to occur in principle,
this generic presence of ramification points leads us 
here, in contrast to a similar case\footnote{We remark that in the sect.~\ref{twist class} where
we investigate a similar example for 'new' classes on $C$, arising from components
of the curve $\si|_C$ which becomes reducible for special values of the moduli,
the situation is at first somewhat similar: in both cases the question whether a 
reducibility of the intersection of $C$ with a surface (here $\E_c$, there $\si$)
is dangerous for the smoothness of $C$ is considered. Although there again in principle 
a harmless interpretation of the potential intersections is possible in analogy
with what we have here, the expectation that these points are 'ramification-like' is
much less standard there; so we will adopt the (highly-restrictive, as it turns out)
condition $\D\cdot \D'=0$ in that latter case.} 
in sect.~\ref{twist class}, 
to adopt the strategy not to demand in addition that $\C_1\cdot \C_2=0$.

\subsection{\label{new class}Why the component $\C_1$ of $\C$ represents a 'new' class}

Let us now investigate whether the class (of the curve) $\C_1$ on $C$,
which according to its definition at least looks different from 
the generically available classes $\si|_C$ and $\pi_C^*\phi$,
is actually 'new', i.e.~not contained in the span of these 'standard' classes.

For this let us assume that one would have a relation in cohomology
(where $k\in {\bf Z}$)
\beqa
\label{first span relation}
\C_1&=&k\; \si|_C + \pi_C^* \phi
\eeqa
The ensuing relation in $H^2(B, {\bf Z})$, which results from the projection
$\pi_{C*}$, would then be 
\beqa
3c&=&kA_B+5\phi
\eeqa
The class $3c\! -\! k\bar{\eta}$, however, 
will not in general\footnote{if the discrete parameters in $\eta$
are not chosen in such a way that $3c\! -\! k\bar{\eta}\!\equiv \! 0 (5)$ for some 
$k\!\in \!\{0,1,2,3,4\}$} 
(the precise conditions have to be considered case by case)
be divisible by $5$ 
(for any $k$, assuming that not the class of $c$ itself is already divisible by $5$),
giving the sought-after contradiction 
(similarly for $n = 4$).

\subsection{\label{reduction counting}The question of moduli reduction}

So if one restricts the bundle moduli (the degrees of freedom coming from the $a_i$) 
by posing along $c\! \cong \! {\bf P^1}$ the condition (\ref{factorization relation over c}), 
one gets the factorization of the equation (\ref{factorization over c}) for $\C$ 
and thus the decomposition (\ref{vertical decomposition}) 
which defines the 'new' cohomology class of $\C_1$.
Asking conversely which moduli restriction is enforced
by demanding the existence of this cohomology class (because it is used in 
a discrete twist) one meets the following problem: first what one really
uses in the twist construction is a line bundle, thus a divisor {\em class} on $C$; so
one has to make sure that an {\em effective} representative in this class exists;
in a second step one has to clarify whether the existence of such a curve
(which we hope to play the role of $\C_1$)
can arise {\em only} in the way (\ref{vertical decomposition})
or whether it may exist 'accidentally' already on a larger moduli subspace 
than the one given by (\ref{factorization relation over c})
(where it exists 'naturally').

Let us consider the question on the numbers of degrees of freedom
in the general versus the factorised case. To keep things simple 
in this illustrating example we did assume that $c\! \cong \! {\bf P^1}$. $\!$Then one gets
as number of parameters of the general equation $w|_{\E_c}\!\! =\!\! 0$ the sum of parameters in the 
homogeneous polynomials $A_i$ of degree $e\! -\! ir$ (where $e:=\eta\cdot c$ and $r:=c_1\cdot c$
and we also assume here that $e,r\geq 0$), so one gets in total $5e-14r+5-1$. 
On the other hand we have in the factorised case the degrees $\deg F_1 = e-2r-E, \deg G_1 = e-4r-E,
\deg H_1 = e-5r-E, \deg F_2 = 2r+E, \deg G_2 =: E$ (with $0\leq E\leq e-5r$ because of $A_5 = H_1G_2$), 
so $3e-9r-E+5-1$ parameters in total.

Let $V_A$ and $V_F$ be the vector spaces generated by the coefficients
of the homogeneous polynomials $A_i$ and $F_1, G_1, H_1, F_2, G_2$, respectively. 
Then (\ref{factorization over c}) gives a (non-linear) map
\beqa
p: V_F&\ra &V_A
\eeqa
As we are interested actually only in the zero divisor of $w|_{\E_c}$ we have to subtract 
in both cases above one ineffective degree of freedom.

Now the degree of the condition (\ref{factorization relation over c}) 
is $3e-10r$, thus the vanishing poses actually $3e-10r+1$ conditions. So when one demands
this condition of the original number $5e-14r+4$ of free parameters only $2e-4r+3$ remain
and one is restricted to a linear subspace 
(or to the corresponding projective subspace)
\beqa
\label{subset relation}
U_A&\subset &V_A
\eeqa
Above, in the paragraph after (\ref{factorization relation over c}),
we investigated the question whether the concrete factorization (\ref{factorization over c})
is even more special than what the condition (\ref{factorization relation over c}) demands 
or whether the latter condition is also already sufficient (and thus equivalent) 
to imply the special form (\ref{factorization over c}), i.e.~whether
the image $\mbox{im}\,  p$ of $p$ is or is not a {\em proper} subset of $U_A$ 
\beqa
\label{coefficients subset relation}
\mbox{im}\, p & \subset & U_A
\eeqa
The comparison of the number $\dim V_F-1$ 
of free parameters in the special from (\ref{factorization over c}) 
with the number $\dim U_A-1$ of parameters left free after posing condition
(\ref{factorization relation over c}) gives
\beqa
\dim V_F -1&=&2e-4r+3+(e-5r-E)+1\\
\dim U_A -1&=&2e-4r+3
\eeqa
(note $E\leq e-5r$).
The answer $\mbox{im}\, p=U_A$ to the mentioned question
(given after (\ref{factorization relation over c}))
should be read combined with the concrete computations of the numbers of degrees of freedom:
the {\em specialising subset} has codimension $3e-10r+1$ in the moduli space.\footnote{The problem, 
alluded to earlier, remains however:
whether not perhaps the divisor class of $\C_i$ ($i=1,2$) exists accidentally on $C$ already 
along a {\em larger subset} of the moduli space.}

A further important issue, especially in connection with the question discussed above 
immediately before sect.~\ref{new class} of whether we have to demand that $\C_1\cdot \C_2=0$ 
or not, is the question whether an irreducible member of the linear system
$|\C|$ exists at all (to see the moduli reduction effect when demanding the reducibility);
the analogous condition $\D\cdot \D'=0$ in sect.~\ref{twist class} 
(to which we referred also in the discussion above which we just mentioned) 
will obstruct just this\footnote{so there will be no question
concerning the codimension of a specialising subset of the moduli space where the reducible
decomposition of a certain curve exists (to pose the cohomological 
condition for the possibility, on a moduli subset, of an {\em orthogonal} decomposition 
is itself a choice between different components of the moduli space 
and not an example of moduli reduction in a given connected component)}
(cf.~the final paragraph of sect.~\ref{moduli reduction}).

{\em Remark:}
The considerations which follow (not used elsewhere) in the rest of this subsection  
are best appreciated after having made acquaintance with the similar arguments 
in the final paragraph of sect.~\ref{moduli reduction} 
and can be easily postponed in a first reading.

In the present section we decided not to pose this orthogonality condition of the components;
but even if one would do so
here the corresponding argument starting from the possible assumption $\C_1\cdot \C_2=0$
(on a subset of the moduli space where $\C$ has become reducible)
would not preclude the existence of an irreducible $\C$.
One may contrast this with the mentioned final paragraph of sect.~\ref{moduli reduction}:
there, if $A_B=C\cap B\subset B$ degenerates to become reducible $A_B=D+D'$, 
the varying moduli in question concern the shape of $C$; $B$, however, is not changed, and
the possibility of the mentioned degeneration shows the existence of the divisor $D$ on $B$
(this existence as such is independent of the specific moduli chosen for $C$),
so it will always (independently of the moduli chosen for $C$) 
make sense to build the intersection product $A_B\cdot D$ in $B$; from this starting point
a contradiction is derived in the final paragraph of sect.~\ref{moduli reduction} which forbids
the existence of an irreducible $A_B$ (under the assumption that an orthogonal decomposition exists).

In the example in sect.~\ref{first example} the situation is different.
If an irreducible $\C$ exists one would have to know the following if one wants 
to derive a potential contradiction (which would forbid, regrettably,
in the end the existence of an irreducible $\C$, given the existence of an orthogonal decomposition):
one would like to argue that $\C\cdot \C_1\geq 0$ from the fact that this 
can be interpreted as a set-theoretic intersection (with no self-intersections involved). 
But here (different from the case on one and the same surface $B$
on which various curves are considered from various moduli choices for $C$)
the components $\C_i$ and a potential irreducible $\C$ do not exist
on one and the same surface $C$ (for a specific moduli choice); 
the product $\C\cdot \C_1$ of cohomology classes can therefore not be
related to the corresponding irreducible divisors (which would imply the non-negativity; 
cf.~also footn.~\ref{only topological cycle}).

This shows a further difference between potential orthogonal decompositions
of the two 'standard' curves, $\C=\E_c|_C=\C_1+\C_2$ here 
and $\si|_C=\D+\D'$ in sect.~\ref{twist class} (they arise as intersections
with either the elliptic surface $\E_c=\pi^{-1}(c)$ or $B$): for $B$
one gets a contradiction assuming an irreducible $A_C=A_B$ because
the curves involved ($A_B, D, D'$) lie all in $B$ and thus exist independently of the
specific moduli chosen for $C$. So one can build $A_B\cdot D$
as the curves exist simultaneously and derive a contradiction from that.

So, in contrast to the case in sect.~\ref{twist class} where we have reasons (as described in the 
final paragraph before sect.~\ref{new class})
to adopt the the orthogonality assumption and where it leads to dramatic restrictions
(among them the nonexistence of an irreducible $A_C$), 
in our present example it does not forbid in principle the existence of an irreducible $\C$.

\newpage

\subsection{\label{case n=4}The case $n=4$}

Finally we consider the other phenomenologically relevant case of $n=4$.
Here one has $A_5=0=H_1$ and thus one gets immediately also $A_3=0$ 
(this was equ.~(\ref{factorization relation over c}) in the case $n=5$) as a 
necessary condition for the factorization 
\beqa
\label{n=4 factorization}
A_0z^2+A_2xz+A_4x^2&=&(F_1z+G_1x)(F_2z+G_2x)
\eeqa

Again we ask whether this condition is also already sufficient.
But the demand that the relevant number of degrees of freedom
contained in the coefficients of the $F_i, G_i$ is at least as large as the corresponding
number for the $A_i$ leads to the inequality $2e-4r+4-1\geq 3e-6r+3-1$ or $\deg A_2=e-2r\leq 1$,
which contradicts $\deg A_4=e-4r$ 
(for $r=c_1\cdot c >0$ which itself is certainly the case if $c_1$ is ample, say).
Thus a further relation between the polynomials $A_0, A_2, A_4$ is needed which
reduces their collective number of degrees of freedom by $e-2r-1$. Of course just such 
a relation follows from (\ref{n=4 factorization}) as one gets from the relations
$A_0=F_1F_2, A_2=F_1G_2 + G_1F_2, A_4=G_1G_2$ the condition 
\beqa
\label{n=4 condition}
A_2^2-4A_0A_4&=&R^2
\eeqa
(with $R=F_1G_2 - G_1F_2$). As the count of the reduced number of the degrees of freedom
contained then in the $A_i$ already suggests this necessary condition is now also sufficient:
(\ref{n=4 condition}) gives $A_4|(A_2+R)(A_2-R)$ and one can write
$A_4=G_1G_2$ with $A_2+R=2F_1G_2$ and $A_2-R=2G_1F_2$ and all the polynomials $F_i,G_i$ 
are now known (up to an overall constant) from the $A_4, A_2$ and $R$; 
taking again into account (\ref{n=4 condition}) shows that the relation $A_0=F_1F_2$ is also 
fulfilled. 
Note that of the a priori possible number $2e-4r+1-1$ of degrees of freedom on the left hand side of 
(\ref{n=4 condition}) only $e-2r+1=\deg R +1$ remain.

So for $n=4$ one has the two equations (which are together necessary and sufficient)
for the factorization: $A_3=0$ and (\ref{n=4 condition}). This poses a number of
$e-3r+1+e-2r-1$ conditions, i.e.~to have the indicated decomposition of $\C=\pi_C^{-1}(c)$
one restricts to a subspace of codimension $(2\eta-5c_1)c$. It remains, of course, 
the known problem of whether a class $\C_i$ of the components does not exist perhaps 
'accidentally' already on a larger subset of the moduli space; on the indicated subspace
the class exists naturally.

The analogue of the relation (\ref{trivial relation}) is here
(whence in particular $\C_i\cdot \si|_C=(g_i)\cdot c$)
\beqa
\label{n=4 trivial relation}
A_B&=&\bar{\eta}=(a_4)=(g_1)+(g_2)
\eeqa
Furthermore one has for the divisors on $\E_c$
\beqa
\C_1&=&\Big(2\si + \eta - 2c_1 -(g_2)\Big)\Big|_{\E_c}\\
\C_2&=&\Big(2\si+2c_1+(g_2)\Big)\Big|_{\E_c}
\eeqa
Thus one gets (as intersection number in $\E_c$ and also\footnote{assuming 
that $\C_1$ and $\C_2$ have no common component 
such that no self-intersection number is involved} in $C$)
\beqa
\label{n=4 intersection evaluation}
\C_1\cdot \C_2 &=& \Big( 2 \bar{\eta}+4c_1\Big) c
\eeqa
Proceeding as in the case $n=5$ one gets with
$\pi_{C*}(\C_i^2)=(2c-2 \bar{\eta}-4c_1)c$ finally ($i=1,2$)
\beqa
\label{first new c_2 n=4}
c_2(V)&=&\eta \si - \frac{5}{2}c_1^2-\frac{1}{2} \eta \bar{\eta}+
4\la^2  \Bigg[\frac{1}{2}\eta\bar{\eta} 
+ \Bigg(2\bar{\eta}-4(g_i)-2c+2\Big( 2\bar{\eta}+4c_1\Big)\Bigg)c\Bigg]
\;\;\;\;\;\;\;\;\;\\
\label{first new Ngen n=4}
-N_{gen}&=&\la \Big[\eta \bar{\eta}+ \Big(2\bar{\eta}-4(g_i)\Big)c\Big]
\eeqa
Here, again, taking $i=1$ or $2$ changes by (\ref{n=4 trivial relation}) 
just the sign of the new term in $N_{gen}$.

Taking into account the parity considerations (cf.~footn.~\ref{integrality footnote}) 
is much easier in our case of $n=4$ here than it was previously for $n=5$
because no parity issue on $C$ is involved as all parity conditions are formulated on $B$;
furthermore the question is even completely independent of the new twist class $\C_i$.
Now $\la$ can be integral or strictly half-integral: in the first case
one has just to demand that $\eta\equiv c_1 \, (2)$ on $B$ 
(or equivalently $\bar{\eta}\equiv c_1 \, (2)$); for strictly half-integral $\la$
one gets the condition that $c_1$ has to be even (as $\pi_{C*}(\C_i)=2c$).

\subsubsection{\label{Concrete examples n=4}Some concrete examples}

We take now $n=4$ and note that $\eta$ and $\eta-4c_1$ have to be effective
(classes of effective divisors), the linear system $\eta$ has to be base point free,
one has the parity condition $\eta\equiv c_1 \, (2)$ and $0\equiv c_1 \, (2)$ for 
$\la$ being integral and half-integral, respectively; further $c\cong {\bf P^1}$ and 
$\deg G_2=E$ has to fulfil $E\leq e-4r=(\eta-4c_1)c$.

We take first, as \un{case $1$}, $B={\bf P^2}$ where $\eta=al$ (with the class $l$ of the line $l$)
gives the conditions $a\geq 12$, $\la\in {\bf Z}$ and $a$ odd.
We take $c=l$, get the condition $E\leq a-12$ and 
\beqa
\label{Ngen n=4 case 1}
-N_{gen}&=&\la \Big[ a(a-12) + 2(a-12)-4E\Big]
\eeqa
($c_2(V)$ is computed similarly). The flexibility from $E$ is obvious.
Taking instead $c=2l$ one gets the condition $E \leq 2a-24$ and the new terms are multiplied by $2$.

We take $B={\bf F_0}$ (with base $b$ and fibre $f$) 
as \un{case $2$} where $\eta=xb+yf$ has to fulfil $x,y\geq 8$
and $x,y$ even for $\la\in {\bf Z}$ (or no further restriction for 
$\la\in \frac{1}{2}+{\bf Z}$). We take $c=f$,
get the condition $E = (g_2)\cdot c=x_g\leq x-8$ (using the notation $(g_2)=x_gb+ y_gf$) and
\beqa
\label{Ngen n=4 case 2}
-N_{gen}&=&\la \Big[ x(y-8) + y(x-8)+ 2(x-8)-4E\Big]
\eeqa
Note that here the (easily won) examples serve just the purpose of mere illustration. 
By contrast in sect.~\ref{twist class},
where we adopt the highly restrictive condition $\D\cdot \D'\! =\! 0$ for the components
of $\si|_C$, they give existence proofs for the non-emptyness of the construction.

\section{\label{twist class}A second example of a non-generic twist class}

\resetcounter

For our next example of a 'non-generic' class $\chi$
we again have to go to a sublocus of the moduli space $\M_V$ where
a twist exists which is not available generically
(but cf.~the discussion in sect.~\ref{moduli reduction}).
We consider the discrete parameters $n, \eta, \la$ fixed
and concentrate just on the connected component $|C|={\bf P}H^0(X, \cO(C))$
of $\M_V$. This is parametrised by the possible different shapes of $C$
lying in $X$; equivalently by the possible different forms of its
defining equation $w=0$ (up to constant rescaling) in $X$
(variations in $\M_V$ are variations in the coefficients $a_i$ of $w$, 
up to an overall multiplicative constant).

The sublocus we are interested in is defined by assuming 
that the equation $w=0$ has a special form:
we assume that the highest coefficient factorises 
(nontrivially: neither $d$ nor $d'$ is a constant)
\beqa
\label{factorization}
a_n &=& d \cdot d'
\eeqa
This has the consequence that its vanishing locus $(a_n)$,
the curve $A_B:= C\cap B \subset B$ of cohomology class $\bar{\eta}$,
becomes reducible
(where $D=(d)$ and $D'=(d')$ 
are curves\footnote{\label{identification}we will often use the same symbol 
for the curves and their cohomology classes; for $(a_n)=A_B$ cf.~[\ref{FMW}]} in $B$)
\beqa
\label{decomposition}
A_B &=& D + D'
\eeqa
Conversely, having such a decomposition into two curves, 
is equivalent\footnote{as $D$ and $D'$ represent
effective divisors, such that sections of the corresponding line bundles
lead back to the factors in (\ref{factorization})}
to the factorization (\ref{factorization}).
Among various such decompositions which exist we consider the one
in (\ref{decomposition}), involving the curves $D$ and $D'$, varying in their
respective linear system. When
we consider in a moment the corresponding decomposition of $\si|_C$ in $C$
we will be interested (as in the end we want to use the twist
by the corresponding line bundle on $C$) only in the divisor {\em class}
of the 'new' component $\D$ which occurs then on $C$
($\D$ is just $D$, considered as curve in $C$); therefore we take into account, 
already in the consideration on $B$, just the divisor {\em class} $[D]$
of $D$ (on $B$ this is equivalent to fix just the 
cohomology class $\delta$ of $D$ in the corresponding 
cohomological decomposition $\bar{\eta}=\delta+\delta'$). Furthermore, with
an eye on the corresponding situation on $C$, we define the 
following two subsets of the moduli space $\M_V$: 
first $\cS_{[D]}$, the subset where the indicated divisor class exists - 
but this, obviously, turns out to be the full moduli space $\M_V$
(here we assume $[D]$, as $[D']$, to be just an effective divisor) - , 
and secondly $\cS_{A_B=D+D'}$, 
which we define as the subspace of $\M_V$ specified by the subspace of $|(a_n)|$
where the curves of divisor class $[A_B]=[(a_n)]$ 
decompose into two curves of the indicated divisor classes.

The same decomposition as in (\ref{decomposition}) does then hold
(equivalently) for the identical point set considered as
curve in the surface $C$, i.e.~for the curve $A_C:=\si|_C=B\cap C\subset C$
(where $\D$ and $\D'$ are just $D$ and $D'$, 
but now considered as curves in $C$)
\beqa
\label{decomposition on C}
A_C &=& \D + \D'
\eeqa

Note that on $B$ the curve $D=(d)$ 
will of course always exist {\em as such} (so it is the {\em decomposition}
(\ref{decomposition}) which 
is equivalent to the factorization (\ref{factorization}));
by contrast, on $C$ already the {\em existence} of the curve $\D$ 
can not\footnote{the divisor class 
of the pullback $\pi_C^* D$ does not equal $\D$ as, even on $\cS_{\D}$, 
$\D$ will be only a component of the pullback $\pi_C^* D$} 
be assumed generically; again we will be interested (for the indicated reasons
of twisting) just in the existence of the divisor {\em class} $[\D]$ of a
specific curve $\D$; this will exist (on $C$) only for a subset $\cS_{[\D]}$ of $\M_V$
(this subset will now be nontrivial in general, in contrast to the situation on $B$). 
When using the twist by $\cO_C(\D)$
one restricts the moduli space from $\M_V$ to $\cS_{[\D]}$.
We will also consider again the subspace $\cS_{A_C=\D+\D'}$ where a
decomposition of the concrete curve $A_C$ into two curves with the indicated 
divisor classes holds on $C$.
Obviously one has 
$\cS_{A_B=D+D'}=\cS_{A_C=\D+\D'}\subset \cS_{[\D]}$.

To be in a region of parameters where $C$ is non-singular 
one has to avoid at least obvious self-intersections.
This leads one to demand 
\beqa
D\cdot D'&=&0
\eeqa 
This turns out to be quite a restrictive condition; below we will treat further the question 
whether such an {\em orthogonal} decomposition (as we will call it) can be assumed to exist.

We assume that we are in the generic case where $D$ and $D'(\neq D)$
do not have a common component (they may be irreducible, for example);
then their intersection number really counts a number of points and one has
$D\cdot D' = \D \cdot \D'$.
By contrast the self-intersection number\footnote{the self-intersection number 
does not just count a number of points 
(which can be seen just from the set-theoretic intersection, adjusted with multiplicities), 
but rather is the degree of the normal bundle}
is sensitive to the ambient surface 
in which the curve is considered to lie: one has $D^2 \neq \D^2$ in general
(cf.~the discussion at the end of sect.~\ref{moduli reduction}).\\
\indent
Similar remarks apply also to the curve $A_B$ in $B$ versus the curve $A_C$ in 
$C$: whereas the first often is assumed to be ample
(though we will not do so, cf.~sect.~\ref{Concrete examples}), 
implying a positive self-intersection number, the latter has  
-  again under mild assumptions, cf.~below  -  
negative self-intersection number, so it is isolated on $C$
(by contrast the linear system $|A_B|$ comprises, as said, easily a continuous
family of equivalent divisors in $B$), all this despite the fact that the
same point set is concerned. Besides the sufficient difference that this 
point set is, as remarked, considered as curve in different surfaces ($B$ and
$C$, respectively), one should also note the different meaning of the issue
of 'movability' in $B$ versus $C$: the movability in $B$ means that 
{\em different} surfaces $C$ (when $C$ varies in its own linear system in $X$)
cut out different curves $A_B=C\cap B\subset B$; by contrast the isolatedness
of $A_C$ refers, of course, to a {\em fixed} surface $C$
(this whole discussion can be carried through analogously also for $D\subset B$ 
versus $\D\subset C$).

Furthermore, as mentioned earlier, one has$^{\ref{identification}}$ 
$\pi_{C*}A_C=\bar{\eta}=A_B$. 
More precisely one has even 
the corresponding relations for the individual components
\beqa
\label{projections}
\pi_{C*}\D \; = \; D , \;\;\;\;\; \pi_{C*}\D' \; = \; D'
\eeqa
Similarly as for $A_C$ in each case the only effect here
of the projection $\pi_C: C\ra B$ is to reinterpret
the relevant curve in $C$ (which lies in the intersection $B\cap C$) as a
curve in $B$.
Furthermore $\pi_{C*}\D^2=\D^2$, {\em understood as numbers},
as will be checked below explicitly.

To compute the contributions in (\ref{c2 contribution}) and (\ref{Ngen contribution}) 
for our case $\chi=\D$ it remains to compute $\pi_{C*} \D^2$ (we do not use $D\cdot D'=0$).
To compute $\D^2$ we compare the canonical class,
considered as a number (the negative Euler number),
of $D\subset B$ and $\D\subset C$
\beqa
\label{canonical in B}
K_D    &=& (K_B + D)|_D    = (D-c_1)D\\
\label{canonical in C}
K_{\D} &=& (K_C +\D)|_{\D} = (n\si|_C + \pi_C^*\eta + \D)|_{\D} 
= (n+1)\D^2+nD D'+\eta D
\eeqa
(using the cohomological relation $D+D'=\eta-n c_1$)
which leads to the relation of numbers 
\beqa
\label{self-intersection on C}
\D^2 &=& \frac{1}{n+1} ( D-\eta-c_1-nD')D \, = \, - c_1 D-DD'
\eeqa
(here in (\ref{canonical in C}) we made use of the relation of numbers
$\pi_C^* \eta \cdot \D = \pi_{C*}(\pi_C^* \eta \cdot \D)=\eta\cdot D$).

Let us check also the relation of numbers $\pi_{C*}\D^2=\D^2$.
Note first that $A_C^2=\si^2|_C=-c_1A_C=-c_1(\D+\D')$ 
(where $c_1$ is actually $\pi_C^* c_1$) is also $\D^2+\D'^2+2\D\D'$;
furthermore one has even individually\footnote{compare 
(\ref{canonical in B}) and (\ref{canonical in C}):~the former
gives $K_D=(\bar{\eta}-c_1-D')D=(\bar{\eta}-c_1-\D')\D=(\eta-(n+1)c_1-\D')\D$ 
as numbers, and the latter $K_{\D}=(\eta+(n+1)\D+n\D')\D$ (always with suitable $\pi_C$ pull-backs)}
$\D^2 = -c_1 \D-\D\D'$ and similarly for $\D'$.
So one gets with the projection formula and (\ref{projections}) that 
$\pi_{C*}\D^2=-c_1\pi_{C*}\D-DD'=-c_1 D-DD'=\D^2$.

Thus one gets from (\ref{c2 contribution}), (\ref{Ngen contribution}) 
in this example of $\chi=\D$ (as cohomology class) the complete expressions
(the first terms in the $[ ... ]$ brackets are the standard 
contributions, the terms proportional to $D$ are the new contributions; we did not yet assume $DD'=0$)
\beqa
\label{new c_2}
c_2(V)&=&\eta \si - \frac{n^3-n}{24}c_1^2-\frac{n}{8} \eta \bar{\eta}+
\frac{n}{2}\la^2 \eta \Big[\bar{\eta} + 3 D\Big]\\
\label{new Ngen}
-N_{gen}&=&\la \eta \Big[\bar{\eta}+D\Big]
\eeqa

\subsection{Why the component $\D$ of $A_C$ represents a 'new' class}

Let us now investigate whether the class (of the curve) $\D$ on $C$,
which according to its definition at least looks different from 
the generically available classes $\si|_C$ and $\pi_C^*\phi$,
is actually 'new', i.e.~not contained in the span of these 'standard' classes.

For this let us assume that one would have a relation in cohomology
(where $k\in {\bf Z}$)
\beqa
\label{span relation}
\D&=&k\; \si|_C + \pi_C^* \phi
\eeqa
The ensuing relation in $H^2(B, {\bf Z})$, which results from the projection
$\pi_{C*}$, would then be
\beqa
D&=&k\; A_B + n \phi
\eeqa
In other words one would get that
\beqa
\label{divisibility demand}
k(D+D')-D&=&n (-\phi)
\eeqa

Here, however, the class $(k-1)D+kD'$ on the left hand side 
will not in general be divisible by $n$,
giving the sought-after contradiction. 

Of course, it is possible that such a divisibility does hold under special circumstances,
for example\footnote{more generally one has to forbid that $D\equiv k\bar{\eta}\, (n)$ 
for any $k=0, \dots, n-1$} 
when one of the classes involved is itself already divisible by $n$:
if one would have, say, $D=n\bar{D}$ one can just take $k=0$ 
in the resulting expression $(k-1)n\bar{D}+kD'$ (and analogously for $D'$). 
In general, however, the demand of divisibility by $n$ of the
left hand side of (\ref{divisibility demand}) poses a condition which a priori
need not to be fulfilled. 
So a relation (\ref{span relation}), which would show that the class
$\D$ on $C$ is not 'new', will not hold in general.

This result looks very promising in what concerns the question of moduli 
reduction by using a twist involving the 'new' class (or rather the 
corresponding line bundle). However there is still another annoying 
possibility which can not be excluded. Specialising the bundle moduli
(actually here the equation of $C$) appropriately one may be able to 
select a locus where $A_B$ decomposes as described; twisting with $\cO(\D)$
will then not be available generically (though the locus of availability
might be larger then it seems at first sight because the class might, for
some reasons, exist already on a somewhat larger subset of the moduli space).
However, when posing the orthogonality condition $D\cdot D'=0$ for the 
components $D$ and $D'$ of $A_B$ one has already posed a cohomological, 
i.e.~{\em discrete} condition. One has not excluded the possibility that
this discrete condition forces the relevant part of the moduli space $|C|$ of deformations of $C$
(inside $X$), i.e.~in our case the moduli space $|A_B|$ of deformations of $A_B$ (in $B$) 
already to decompose into the corresponding moduli spaces
of the deformations (in $B$) of $D$ and $D'$. In other words, the 
discrete condition might enforce already that no irreducible member of $|A_B|$
exists.

\subsection{\label{moduli reduction}The question of moduli reduction}

Note first that a reducibility of one 
of the generically known classes\footnote{i.e.~the classes consisting of 
$\si=\si|_C$ and the pull-back classes $\pi_C^* \phi$ for 
corresponding classes $\phi$ on $B$} 
on $C$ (in our case here the class $\si=\si|_C=A_C$)
does, a priori, not necessarily always introduce a 'new' class 
(linearly independent of the classes which are already present generically). 
For example, one has already generically the reducible
decomposition $\pi_C^* A_B=A_C + \tilde{A}_C$ for some further class
$\tilde{A}_C$ which is however not 'new' as it equals $\pi_C^* A_B - \si|_C$.

As already remarked above, when using the twist by $\cO_C(\D)$
one restricts the moduli space from $\M_V$ to $\cS_{[\D]}$
(the existence of the line bundle is equivalent to the existence of the 
divisor class).
Now, a concrete description of the stabilized subspace $\cS_{[\D]}$ 
is less immediate than in the completely 
explicit\footnote{because this subspace refers directly to a
specifying condition on $a_n$, cf.~(\ref{factorization}), 
and the $a_i$ directly describe the moduli space $\M_V$} 
case of $\cS_{A_B=D+D'}$: the latter is, however, in general 
only a subspace of the former: $\cS_{A_B=D+D'}\subset \cS_{[\D]}$.

To bring these two subspaces in a useful relation, i.e.~to relate 
the stabilized subset $\cS_{[\D]}$ to the explicitly
describable subset $\cS_{A_B=D+D'}=\cS_{A_C=\D+\D'}$, poses however the problem
to go from the mere existence of the divisor {\em class} $[\D]$ to the existence
of a member of it, here a concrete {\em effective} divisor $\tilde{\D}$, say,
which furthermore should then constitute a component of $\si|_C$
(such that the curve $\si|_C$ decomposes as $\si|_C=\tilde{\D}
+\tilde{\D'}$, cf.~(\ref{decomposition on C})).

Note that although $\D$ is by definition a component
of $\si|_C$ (where the latter decomposes only along a certain subset of $\M_V$)
one can not exclude the possibility that, firstly, the curve $\D$ as such
exists\footnote{this refers to a curve on each surface $C$ corresponding to a
point in a part $\cS_{\D}$
of the moduli space ${\bf P}H^0(X, \cO(C))$, which comprises but is larger than
the subset $\cS_{A_C=\D+\D'}$, and which specialises  -  when going to 
the latter subset of the moduli space  -  to the component of $\si|_C$ which
carries the name $\D$} 
on some surfaces $C$ without $\si|_C$ being reducible (with $\D$ as a component), 
and, secondly, the divisor {\em class} 
of the curve $\D$ may exist on an even greater subspace $\cS_{[\D]}$ 
of $\M_V$, i.e.~the problem has actually two parts: a priori
one knows only $\cS_{A_B=D+D'}=\cS_{A_C=\D+\D'}\subset
\cS_{\D}\subset \cS_{[\D]}$ where we denote by $\cS_{\D}$ the subspace of 
$\cS_{[\D]}$ where an effective member (i.e.~a real curve) exists.
Here the inclusions are in general not equalities and reflect the different
steps of the problem referred to before; we will consider them respectively below.
Both steps are not easily controlled (i.e.~specialising conditions which make
both inclusions equalities are not easily provided). 
So in this example of $\chi=\D$, where
we can compute quite explicitly new contributions to the chiral matter,
it is not straightforward to describe, when the twisting with 
$\cO_C(\D)$ restricts the moduli from $\M_V$ to $\cS_{[\D]}$, 
how much the latter is larger than the 'known' subset $\cS_{A_B=D+D'}$
(which has an explicit description as a moduli space subset, 
cf.~(\ref{factorization})).

In the {\em first step} (to go from $\cS_{[\D]}$ to $\cS_{\D}$) one has 
to secure the existence of an {\em effective} member $\tilde{\D}$
in the linear equivalence class $[\D]$ (i.e.~$|\D|\neq \emptyset$ or $H^0(C, \cO_C(\D))\neq 0$); 
although $[\D]$ is by definition the {\em class} of an effective
divisor ($\D$, which occurs as component of $A_C$ {\em along a certain subset of $\M_V$}) 
it could happen that $H^0(C, \cO_C(\D))$ 
(where, despite the notation, only the divisor {\em class} of $\D$ is actually present) 
is zero generically on $\cS_{[\D]}$ and jumps upwards only on a proper subset (which is $\cS_{\tilde{\D}}$
with $\tilde{\D}$ an {\em effective} divisor in $[\D]$).

More precisely what can be said is the following.
Assume that a divisor $\F$ on $C$ exists (on $\cS_{[\D]}$)
which, after going to $\cS_{\D}$, 
becomes linearly equivalent to $\D$ 
(which itself exists only after going to $\cS_{\D}$). 
Assume first that $\F$ is {\em effective}: then one gets,
if the linear system $|\F|$ constitutes a {\em continuous} family,
a contradiction as $\D$ can not be moved in any hypothetical family
of linearly equivalent divisors as it has negative 
self-intersection $\D^2=-c_1 \D <0$ when one makes the assumption - as we
do from now on - that $c_1$ is 
ample\footnote{furthermore we assume that $\D$ (or $\D'$) 
and a divisor representing $c_1$ (or rather $\pi_C^* c_1$) 
do not have a component in common: the intersection
number $c_1\cdot \D:=\pi_C^* c_1 \cdot \D$ (in $C$)
counts then really a (weighted) number of points 
and equals $c_1\cdot D>0$ (in $B$) as $c_1$ is assumed to be ample; 
we will also assume that $\D$ is irreducible:
this assumption implies also that a hypothetical 
linearly equivalent divisor $\F$ cannot have a component in common with $\D$
(it is also not possible that $\F$ has $\D$ as component); this assumption
makes sure that the intersection number $\F \D$ is 
really a (weighted) number of points and so non-negative};
if on the other hand $|\F|$ is {\em discrete}
(an isolated $\F$ may of course have, just like $\D$, 
a negative self-intersection number) then note that 
$H^0(C, \cO(\F))\cong H^0(C, \cO(\D))$ is one-dimensional 
(as $\D$ is isolated under our assumptions 
because of $\D^2<0$ and as we assumed $\D$ irreducible) 
and one has $(s)=\D$ just like $(s)=\F$ for a nontrivial
section $s$ of the line bundle which 
depends only on the divisor {\em class}; so such an $\F$ is
actually $\D$ itself and one would get, for an {\em effective} $\F$ 
(and under our assumptions), that $\cS_{[\D]}=\cS_{\D}$.\\
\indent
Now consider, however, the case that $\F=\G-\cH$ is a representation of a 
{\em non-effective} $\F$ as difference of two effective divisors
(actually one can assume that $\G$ and $\cH$ are 
ample\footnote{only the divisor {\em class} 
of $\F$ is important as the relevant property of $\F$
is that it is linearly equivalent to $\D$ (on $\cS_{\D}$); 
$\F$, like any divisor, is linearly equivalent
to the difference of two (very) ample divisors, cf.~Ch.~1, Lemma 5, 
{\em Algebraic Surfaces and Holomorphic Vector Bundles}, 
R. Friedman, (1998) Springer.}).
So, our question is, whether it is possible that $\G$ becomes on $\cS_{\D}$
linearly equivalent to $\D+\cH$; for example, a special case would be 
that it becomes even equal to that combination
(this is somewhat reminiscent of the decomposition $\si|_C=\D+\D'$ along $\cS$
with the decisive difference\footnote{furthermore $A_C$ can not be ample
as this would give $A_C^2=-c_1 A_C <0$} 
that $\D'$ cannot be assumed to exist outside $\cS$).
There are, it seems, no obvious conditions to exclude such a situation, and so
this step leads to an uncontrollable modification ($\cS_{\D}\longleftrightarrow \cS_{[\D]}$) 
of the relevant subset of the moduli space.\\
\indent
In a {\em second step} (to go from $\cS_{\D}$ to $\cS_{A_C=\D+\D'}$)
one must ensure $\si|_C$ decomposes with component $\D\!$ from its mere existence; 
$\!$again there are no obvious conditions ensuring this.

After this general discussion let us bring in now, however,
the orthogonality condition $D\cdot D'\!\!=\!\! 0$. This, unfortunately, excludes
the existence of an irreducible member of $|A_B|$ (so a moduli reduction
effect can actually not be seen in this example; 
we adpot here the mild assumption $H^2(B, {\bf Z})$ 
torsion-free)\footnote{\label{only topological cycle}note that we do {\em not} claim 
a contradiction from trying to argue that $A_C\cdot \D=A_B\cdot D$ 
(as no self-intersection numbers would be concerned which for one curve 
can differ in different ambient surfaces) 
while $A_C\cdot \D=\D^2<0$ (cf.~(\ref{self-intersection on C}), we assumed $c_1$ to be ample) whereas 
$D^2\!\! =\!\! A_B\!\cdot \! D\!\geq \! 0$ 
(as $A_B\!\cdot \! D$ is a weighted set-theoretic intersection);
here (the cohomology class of) the complex curve
$\D$ would persist beyond $\cS_{\D}$ as (the class of) a {\em topological} cycle 
(in the family of surfaces $C$, parametrised by the (connected component of) the moduli space 
$\M_V\!\cong \! {\bf P}H^0(X, \cO_X(C))\! \cong \! |C|$)
which can coexist with (the class of) an {\em irreducible} member of $|A_C|$, 
thus avoiding self-intersections in $A_C\!\cdot \!\D$ 
and making $A_C\!\cdot \!\D\! =\! A_B\!\cdot \! D$ possible a priori; but we do {\em not} argue
that these intersection products would be equal {\em because} they now (without self-intersections) 
would both equal just the weighted set-theoretic intersection: this argument is not at our disposal
as $\D$ persists beyond $\cS_{\D}$ only topologically but not complex analytically (by contrast 
on $B$ members of $|A_B|$ and $|D|$ can coexist, completely independently of the moduli
chosen for $C$, as irreducible {\em complex} curves thus giving $A_B\cdot D\geq 0$); 
we do not claim $A_C\cdot \D=A_B\cdot D$}:~$\!$for in this case, 
with a hypothetical {\em irreducible} curve $A_B$, 
$0\leq A_B\cdot D + A_B\cdot D'=A_B^2\leq 0$ (as either $A_B^2<0$
or $A_B^2\geq 0$ such that $A_B$ is {\em nef}\footnote{this means "numerically effective", 
i.e.~fulfilling $h\cdot c \geq 0$ for all irreducible curves $c$; it implies $h^2\geq 0$},
thus not$^{\ref{Exercise footnote}}$ {\em big} (i.e.~$A_B^2>0$)),
such that $D^2=D'^2=0$ giving a contradiction\footnote{let $H$ be an ample divisor,
$d:=HD, d':=HD'$; then $D'':=d'D-dD'$ ($\neq 0$ adopting the technical assumption $D\neq qD'$ for 
$q$ rational) gives $D''H=0$ but $D''^2=0$ violating the Hodge index theorem}.
This phenomenon will be substantiated in great detail in the explicit examples below.

\subsection{\label{Concrete examples}Concrete examples for the decomposition}

We still have to investigate how restrictive 
our assumption of an {\em orthogonal} decomposition (\ref{decomposition}) is
(we always assume the decomposition {\em non-trivial} , i.e.~$D\neq 0\neq D'$,
and {\em effective}, i.e.~$D$ and $D'$ effective).
Note first that in the spectral cover construction the class
$\bar{\eta}$ of $A_B$ is assumed to be effective; 
often one demands further that it is even {\em ample} 
(i.e.~fulfilling $h^2\! >\! 0$ and $h\cdot c \! >\! 0$ 
for all irreducible curves $c$; 
the individual terms $D$ and $D'$ can in any case not be ample because of the orthogonality). 
But a class $h$ which is ample (or even only big and nef)
is known\footnote{\label{Exercise footnote}Ch.~1, Ex.~13, 
{\em Algebraic Surfaces and Holomorphic Vector Bundles}, 
R. Friedman, (1998) Springer}
{\em not} to admit an orthogonal decomposition. 

So, in searching for a (non-trivial, effective) orthogonal decomposition,
we must assume\footnote{this is something we have to assume for $A_B$;
we know already that $A_C$ is not ample as $A_C\cdot \D <0$}
that the effective class $\bar{\eta}$ is not ample (the argument for the
absence of continuous moduli of the spectral line bundle on $C$ is then
not available, cf.~the remarks at the end of the introduction of 
sect.~\ref{set-up}), and not even big and nef. To make the discussion concrete
we consider the cases $B\! =\! {\bf F_k}$, a Hirzebruch surface, 
or ${\bf dP_k}$, a del Pezzo surface.

\subsubsection{\label{Hirzebruch cases}Examples for $B$ a Hirzebruch surface}

The surface ${\bf F_k}$ is a ${\bf P^1}$-fibration over
a base ${\bf P_1}$ denoted by $b$ (the fibre is denoted by $f$;
as no confusion arises $b$ and $f$ will denote also the 
cohomology classes). One has $c_1({\bf F_k})=2b+(2+k)f$ and the curve
$b$ of $b^2=-k$ is a section of the fibration; 
there is another section ("at infinity") having the
cohomology class $b_{\infty}=b+kf$ and the self-intersection number $+k$; 
note that $b_{\infty}\cdot b = 0$.
A class $(x,y):=xb+yf$ is ample exactly if$^{\ref{Hartshorne}}$ 
$(x,y)\cdot f > 0$ and $(x,y)\cdot b >0$, i.e.~if $x>0, y>kx$.
An irreducible non-singular curve of class $xb+yf$ exists exactly 
if\footnote{\label{Hartshorne}Cf.~Corollary 2.18, Chap.~V, 
{\em Algebraic Geometry}, R. Hartshorne, Springer Verlag (1977).} 
the class lies in the ample cone (generated by the ample classes) 
or is one of the elements $b, f$ or $ab_{\infty}$ (the last only for $k>0$; here $a>0$) 
on the boundary of the cone; these classes
together with their positive linear combinations span the effective cone 
($x,y\geq 0$). $c_1$ is ample for ${\bf F_0}$ and ${\bf F_1}$, whereas for 
${\bf F_2}$, where $c_1=2b_{\infty}$ (such that $c_1\cdot b=0$),
it lies on the boundary of the cone.

Let us now present certain classes $\bar{\eta}=(x,y)$ 
which are effective, but {\em not} numerically effective,
and corresponding orthogonal decompositions of the class $\bar{\eta}$ (of $A_B$): 
on any ${\bf F_k}$ (where $k=0,1,2$)
take $x=0$ and $y\geq 2$ such that (with $y_i>0$ and $y=y_1+y_2$) 
\beqa
\label{first ex}
(0,y)&=&y_1\, f + y_2\, f
\eeqa 
and on ${\bf F_k}$ with 
$k=1$ or $2$ take $y-kx<0$ (and $y>0$), with $y$ even for $k=2$, such that
\beqa
\label{second ex}
(x,y)&=&(x-\frac{y}{k})\, b+\frac{y}{k} \, b_{\infty}
\eeqa 
(${\bf F_2}$ in (\ref{first ex}), (\ref{second ex})
is actually excluded under our assumption that $c_1$ is ample.)\footnote{The twist using $\D$, 
from $D=(0, y_1)$ or $(x-\frac{y}{k}, 0)$, 
needs $y_1$ or $x-\frac{y}{k}$ even for $n$ odd 
and $k+y$ or $x$ and $k+y$ even for $n$ even, $\la \in {\bf Z}$
and $y_1 - k$ or $x-\frac{y}{k}$ and $k$ even for $n$ even, $\la \in \frac{1}{2}+{\bf Z}$
by footn.~\ref{integrality footnote}.}

One gets for the cohomological contributions
from (\ref{first ex}), say, ($F$ the elliptic fibre)
\beqa
c_2(V)&=& \eta \si +\Big[ -\frac{n^3-n}{3}-\frac{1}{4}n^2y+ n^2\la^2(y+3y_1)\Big] F\\
\label{Ngen first F_k ex}
-N_{gen}&=& 2n\la(y+y_1)
\eeqa
(with $\eta=(2n, 2n+nk+y)$ fulfilling the condition $\eta\cdot b \geq 0$).
Here the effect of turning on the new twist is seen directly in a numerical
example: without the twist using $\D$ (from $D=(0, y_1)$)
one would get here only the expressions with $y_1=0$; 
this shows manifestly the greater flexibility achieved by using the twist
(similarly one computes for (\ref{second ex})).

However, although the classes $\bar{\eta}$ in (\ref{first ex}), \ref{second ex}) 
fulfill all the postulated demands
they suffer from another problem: no {\em irreducible} curve realising them
exists\footnote{Actually the same is true for all of their constituents 
on the right hand sides of these equations except for the cases
$y=2$  in (\ref{first ex}) and $x-y/k=1$ in (\ref{second ex})
where $D$ and $D'$ have irreducible representatives.};
so the reduction effect 
(from the set of {\em all} curves of class $\bar{\eta}$, including irreducible ones, 
to those reducible representatives corresponding to a factorisation (\ref{factorization})) 
can not be seen in that case\footnote{\label{no-go footnote}Actually one sees 
from the description given above that irreducible curve representatives exist, 
besides the ample classes which can not be decomposed orthogonally,
only for $b$, $f$ and $ab_{\infty}$ for $k>0$ (i.e.~$k=1$);
but the latter is still simultaneously big and nef 
and the two remaining classes can obviously not be decomposed.
Said differently, {\em on ${\bf F_k}$ a decomposable class} (in our sense)
{\em has no irreducible representative} 
(cf.~for the corresponding situation on ${\bf dP_k}$ the discussion around
(\ref{difference}) below)}.
This (and a similar result which we get below for $B={\bf dP_k}$)
is in the end not too painful as 
with the choice $\D$ (coming from (\ref{decomposition}))
of sect.~\ref{twist class} for the twist class $\chi$ 
the moduli reduction is not under good control anyway, 
as described in sect.~\ref{moduli reduction}. 

\subsubsection{Examples for $B$ a del Pezzo surface}

As second relevant class of base surfaces $B$ let us consider the 
del Pezzo surfaces ${\bf dP_k}$: they are the blow-up of ${\bf P^2}$
at $k$ points $P_i$ for $k=0, \dots, 8$ 
(lying suitably general, i.e.~no three points lie on a line, no six on a conic);
the exceptional curves from these blow-ups are denoted by 
$E_i$, $i=1, \dots, k$ 
(one has ${\bf dP_1}\cong {\bf F_1}$ with $E_1$ corresponding to $b$).
The intersection matrix for $H^{1,1}({\bf dP_k})$ in the basis 
$(l, E_1, \dots, E_k)$, 
with $l$ the proper transform of the line $\tilde{l}$ from ${\bf P^2}$, 
is just $Diag(1, -1, \dots, -1)$; 
furthermore $c_1({\bf dP_k})=3l-\sum_i E_i$ such that $c_1^2({\bf dP_k})=9-k$.
On these surfaces one finds many further examples of orthogonal
decompositions $\bar{\eta}=A_B=D+D'$ 
(of a given curve class into classes of two curves),
among them\footnote{furthermore one can easily 'enhance' a given solution:
take, for example, the one in (\ref{special solution}) with $y_i=1$; from this
one can derive the further solution $2l-2E_1-E_2-E_3 = [ l- E_1-E_2 ] + [ l- E_1-E_3 ]$}
the following families of examples on the first five del Pezzo surfaces 
${\bf dP_k}$, $k=1, \dots, 5$
(with\footnote{note that $l\ra b_{\infty}$ 
and $E_1\ra b$ under ${\bf dP_1}\cong {\bf F_1}$, 
so (\ref{special solution}) corresponds just to (\ref{first ex}) as 
$l-E_1\ra f$} $y=y_1+y_2$ and $y_i> 0$; 
further $b,c\geq 1$ in (\ref{dP_2 solution}) and the parameter $a$ is restricted by 
$1\leq a \leq 2$ in (\ref{dP_3 solution}),
$1\leq a\leq 4$ in (\ref{dP_4 solution})
and $2\leq a \leq 4$ in (\ref{dP_5 solution}))
\beqa
\label{special solution}
yl-yE_1          &=&\Big[ y_1 l-y_1 E_1 \Big] + \Big[y_2 l- y_2 E_1\Big]\\
\label{dP_2 solution}
(b+c+1)l-(b+1)E_1-(c+1)E_2 
&=&\Big[ (b+c) l-bE_1-cE_2\Big]  + \Big[ l- E_1-E_2\Big] \\
\label{dP_3 solution}
(a+2)l-2\sum_{i=1}^2E_i-(2a-1)E_3 &=&
\Big[ al-\sum_{i=1}^2E_i-(2a-2)E_3\Big] 
+\Big[ 2l-\sum_{i=1}^3E_i\Big]\\
\label{dP_4 solution}
(a+2)l-2\sum_{i=1}^2E_i-a\sum_{i=3}^4E_i &=&
\Big[ al-\sum_{i=1}^2E_i-(a-1)\sum_{i=3}^4E_i\Big] 
+\Big[ 2l-\sum_{i=1}^4E_i\Big]\;\;\;\;\;\;\;\;\; \\
\label{dP_5 solution}
(a+2)l-2\sum_{i=1}^4E_i-(2a-3)E_5 &=&
\Big[ al-\sum_{i=1}^4E_i-(2a-4)E_5\Big] +\Big[ 2l-\sum_{i=1}^5E_i\Big] 
\eeqa
That all the classes here are effective\footnote{Furthermore
the condition for base-point freeness of $|\eta|=|\bar{\eta}+nc_1|$, 
cf.~footn.~\ref{base-point footnote},
is easily checked using that ${\bf dP_1}\cong {\bf F_1}$ and that for
$2\leq k \leq 4$ the elements $E_i$ and $l-E_i-E_j$ (where $i\neq j$)
are generators of the effective cone of ${\bf dP_k}$ 
with the relevant properties mentioned in footn.~\ref{base-point footnote}; 
in the example (\ref{dP_5 solution}) on ${\bf dP_5}$ 
one checkes the condition also for the further generator 
(with the relevant properties) $2l-\sum_{i=1}^5 E_i$.}
is seen when reading the class $dl-\sum e_iE_i$ as class of the proper transform
of the corresponding degree $d$ curve in ${\bf P^2}$ which goes 
$e_i$ times through the points $P_i$ and 
noting\footnote{\label{number of conditions}Note 
that $h_d=h^0({\bf P^2}, \cO(d\tilde{l}))$
so one can pose, including multiplicities, $h_d-1$ conditions (of going
through certain points) on these degree $d$ curves. 
Note also that a curve {\em in the original surface} ${\bf P^2}$ 
going through a point $P_i$ with multiplicity $e_i$ 
(what poses $\sum_{j=0}^{e_i-1} j+1 = \sum_{j=1}^{e_i}j$ conditions, cf.~Ex.~5.3, Ch.~I, 
{\em Algebraic Geometry}, R. Hartshorne, Springer Verlag (1977)) would be singular for $e_i>1$;
for {\em its proper transform in the blown-up surface ${\bf dP_k}$ however}
the different local branches going through $P_i$ 
can be separated as they can be choosen in ${\bf P^2}$ to have different slope.} 
that one always has 
\beqa
\label{inequality}
h_d:=\frac{(d+2)(d+1)}{2} > \sum \frac{e_i(e_i+1)}{2}
\eeqa
This amounts to $d+1>e_1$ in the case of ${\bf dP_1}$; here one has to
distinguish the subcases $e_1<d$ and $e_1=d$: in the last case clearly
{\em reducible} realisations of $\sum_{m=1}^d (l-E_1)$ exist.

That the relevant classes have {\em irreducible} curve representatives 
amounts, however,
to demanding something more: the class $2l-2E_1$ on ${\bf dP_1}$, for example,
has effective representatives but these are reducible; 
consider the situation on ${\bf P^2}$ and fix the overall scaling of the
quadratic polynomial by considering the family $X^2+\alpha Y^2+\beta Z^2
+\gamma XY + \delta XZ + \epsilon YZ$; furthermore go to the 
affine $(Z=1)$-patch (i.e.~the $(X,Y)$-plane) and take $P_1=(0,0)$; 
then the demand $e_1=1$ for an ordinary (nonsingular) point
amounts to the one condition $\beta=0$, whereas
a node with $e_1=2$ poses the two further conditions $\delta=\epsilon =0$,
giving three conditions in all; the remaining two-dimensional parameter space
is now, however, already exhausted by the {\em reducible} quadrics, i.e.~the pairs
of lines going through $P_1$, each of them with arbitrary slope (these
reducible quadrics have also already a two-dimensional parameter space); more
explicitly, the ensuing vanishing equation $X^2+\alpha Y^2 + \gamma XY = 0$ 
for the quadric has now the splitting form $(X-aY)(X-bY)=0$. 
This non-existence of an {\em irreducible} representative is in accord 
with the consideration on ${\bf F_1}$: 
an irreducible curve representative for a class $yf$ with $y>1$ does 
{\em not} exist, and these classes correspond just to the classes $yl-yE_1$ 
on ${\bf dP_1}$; 
rather such representatives exist, besides the classes $b,f,ab_{\infty}=a(b+f)$ 
(with $a>0$)
which correspond to $E_1, l-E_1, al$, just for the classes $(x,y)$ with 
$y>x>0$ which correspond to $yl-(y-x)E_1$, 
i.e.~$dl-e_1E_1$ with $0<e_1<d$
(here the boundary cases have been discussed already: 
for $e_1=0$, corresponding to the cases $db_{\infty}$, 
irreducible representatives exist, whereas for $e_1=d$, as described after
(\ref{inequality}), reducible realisations exist).

This shows that the first example (\ref{special solution}) on ${\bf dP_1}$
can not be used for the moduli space reduction argument, 
in accord with the analogous comment on the corresponding case on ${\bf F_1}$ 
at the end of sect.~\ref{Hirzebruch cases}.
This problem is not an accident: rather the phenomenon that here no irreducible curve realisation 
of the class $A_B=\bar{\eta}$ on the left hand side of (\ref{decomposition}), 
or concretely (\ref{special solution})-(\ref{dP_5 solution}), exists
(and rather all realisations are exhausted by following, on the curve level, 
the reducible decomposition on the right hand side) holds in general.
For this assume a decomposition of the mentioned class $A_B=cl-\sum_i e_i E_i$
into the constituents $D=al-\sum f_i E_i$ and $D'=bl-\sum g_i E_i$
being given:
\beqa
(a+b)l-\sum_i (f_i+g_i)E_i &=& \Big[ al-\sum_i f_i E_i \Big] + \Big[ bl-\sum_i g_i E_i \Big]
\eeqa
(i.e.~$c=a+b$ and $e_i=f_i+g_i$).
Then one computes for the difference (between the left and right hand side) 
of the numbers of the available degrees
of freedom 
\beqa
\label{difference}
\frac{(a+b+1)(a+b+2)}{2}-1-\sum \frac{(f_i+g_i)(f_i+g_i+1)}{2}\nonumber\\
- \Bigg[ \frac{(a+1)(a+2)}{2}-1-\sum\frac{f_i(f_i+1)}{2}
       +\frac{(b+1)(b+2)}{2}-1-\sum \frac{g_i(g_i+1)}{2} \Bigg]\nonumber\\
=ab-\sum f_ig_i
\eeqa
The latter expression vanishes now, however, due to the orthogonality 
condition $D\cdot D'=0$ (cf.~for this negative result also the corresponding
situation in footn.~\ref{no-go footnote} for ${\bf F_k}$).

Let us also consider the parity issue (cf.~footn.~\ref{integrality footnote}): using $\D$, 
from $D$ in (\ref{special solution}) - (\ref{dP_5 solution}), needs $y_1, b, c$ even for $n$ odd
and $y$ odd or $b,c$ even for $n$ even, $\la\in {\bf Z}$ and $y_1$ odd for $n$ even, $\la \in 
\frac{1}{2}+{\bf Z}$; other cases, 
in particular (\ref{dP_3 solution}) - (\ref{dP_5 solution}), are excluded.

It is straightforward to evaluate (\ref{new c_2}) and (\ref{new Ngen})
for the different cases. 
As this was done in sect.~\ref{Hirzebruch cases} already for 
(\ref{special solution}) let us just consider
the next infinite series in (\ref{dP_2 solution}): so we take $B={\bf dP_2}$
and have
\beqa
\eta&=&\Big(b+c+1+3n\Big)l-\Big(b+1+n\Big)E_1-\Big(c+1+n\Big)E_2
\eeqa

One computes as result for the chiral matter
(the expression for $c_2(V)$ is complicated)
\beqa
\label{Ngen dP ex}
-N_{gen}&=&\la\Big[ \Big(2bc+2n(b+c)+n-1\Big)+\Big( 2bc+2n(b+c)\Big)\Big]
\eeqa
To realize the enhanced flexibility of our extended ansatz using the 'new' twist class
one should note the following: first of all the second large round brackets 
enclose the 'new' term proportional to $D$, cf.~(\ref{new Ngen})
(had one used $D'$ instead, the new term would be just $n-1$); so although here the same 
parameters $b$ and $c$ occur in the standard contribution and the new contribution
there is a freedom hidden here to have the second part at all.

\section{\label{Conclusions}Conclusions}

An intensely studied class of supersymmetric particle physics models in four dimensions
coming from string theory is that of heterotic models, built from 
a stable holomorphic vector bundle $V$ on a Calabi-Yau threefold $X$.
Two main lines of research are concerned with the particle spectrum [\ref{het spec}],
especially with respect to realistic phenomenology,
and the occurring moduli [\ref{het mod}] and their potential stabilisation.
With regard to the latter the problem concerns 
geometric (K\"ahler and complex structure) moduli from $X$ and bundle moduli. 
As the stabilisation of the latter is a difficult and complex task
it is already interesting to restrict the bundle moduli to a smaller subspace. 
A possibility to achieve this is to make discrete modifications of a given bundle construction 
which are available only over a subset of the bundle moduli space 
such that the twisted bundle has less parametric freedom
(i.e.~turning on such discrete 'twists' constrains the moduli 
which thereby are restricted to a subset of their moduli space)\footnote{We add a word of caution
to exclude possible misunderstandings: when we speak of ``moving to special points in the bundle
moduli space $\M_V$'' to obtain new line bundles on $C$ that can change the topology of $V$
we understand that a corresponding twist is actually made
(the topology of $V$ as such can not change of course); thereby one reaches a {\em new} bundle
$V'$ which has its own moduli space $\M_{V'}$ which is now the subspace of $\M_V$ where the 
twist exists.}.

This idea can be studied concretely in the class 
of spectral cover bundles on elliptically fibered $X$ [\ref{FMW}].
At this point, remarkably, a second highly relevant issue enters the story naturally: 
the non-generic twists lead also to new contributions of chiral matter which modifies
the standard formula [\ref{C}] for the generation number $N_{gen}$
via the appearance of new terms with new parameters. 
This is interesting
as model builders in heterotic string theory have a long, and sometimes woebegone, experience
how restrictive the simultaneous fulfillment of all the phenomenologically relevant conditions is; 
notable among these conditions is the one for $N_{gen}$. 
Seen from this perspective
any method to gain greater flexibility in this class of models is of utmost interest. 
It will be even more welcomed when its use comes 
with the extra bonus of restricting the bundle moduli.

In the present note we develop in sect.~\ref{set-up} 
first the general form (\ref{c2 contribution}), (\ref{Ngen contribution})
of the new contributions to $c_2(V)$ and $c_3(V)$ in the case of the
spectral cover construction (this constitutes a first layer of concreteness)
which are the cohomological quantities relevant for anomaly cancellation condition 
and the generation number, respectively. Then, in sect.~\ref{first example}
and \ref{twist class},  we compute in (\ref{first new c_2}), (\ref{first new Ngen}),
(\ref{first new c_2 n=4}), (\ref{first new Ngen n=4}) and (\ref{new c_2}), (\ref{new Ngen})
everything explicitely 
in the two examples we give for the general type of the needed 'twist class' (second layer).
In both cases it arises from components of a known class (of a curve on the spectral cover surface)
which becomes reducible for a special subset of the bundle moduli space (one problem occurring here
is that, although the mentioned subspace where the new class occurs naturally can be given precisely,
it can not be excluded that the class exists 'accidentally' already on a somewhat larger 
subspace)\footnote{another problem in the second example is 
that the rather strict condition adopted there that the new components are orthogonal 
actually forbids any irreducible representant in the first instance;
this does not constitute any problem, however, for the application to the generation number}.
We also give arguments that generically the classes involved are 'new' in the sense that they
do not belong to the span of the known classes. In both examples we finally specialise even further
and give fully explicit examples for the two different general types of twist class (third layer):
the occurrence in (\ref{Ngen n=4 case 1}), (\ref{Ngen n=4 case 2}) 
and (\ref{Ngen first F_k ex}), (\ref{Ngen dP ex})
of new terms with new parameters in $N_{gen}$ 
shows clearly the enhanced flexibility or the more general ansatz employed. 
This type of procedure should be quite useful for more flexible model building.

I thank the DFG for support in the project CU 191/1-1 and SFB 647 and the FU Berlin for hospitality. 

\vspace{-0.9cm}

\section*{References}
\begin{enumerate}

\vspace{-0.2cm}

\item
\label{FMW}
R. Friedman, J. Morgan and E. Witten, {\em Vector Bundles and F-Theory},
hep-th/9701162, Comm. Math. Phys. {\bf 187} (1997) 679.
 
\item
\label{C}
G.~Curio, {\em Chiral Matter and Transitions in Heterotic String Models},
hep-th/9803224, Phys.Lett. {\bf B435} (1998) 39.

\item
\label{DW}
Ron Donagi, Martijn Wijnholt,
{\em Higgs Bundles and UV Completion in F-Theory},
arXiv:0904.1218.

\item
\label{cxstrfix}
Lara B. Anderson, James Gray, Andre Lukas, Burt Ovrut,
{\em Stabilizing the Complex Structure in Heterotic Calabi-Yau Vacua},
hep-th > arXiv:1010.0255, JHEP 1102:088,2011.\\
Lara B. Anderson, James Gray, Andre Lukas, Burt Ovrut,
{\em The Atiyah Class and Complex Structure Stabilization 
in Heterotic Calabi-Yau Compactifications},
arXiv:1107.5076.

\item
\label{cxstrfix2}
Andreas P. Braun, Andres Collinucci, Roberto Valandro,
{\em G-flux in F-theory and algebraic cycles},
arXiv:1107.5337.

\item
\label{het spec}
Andre Lukas, Burt A. Ovrut, Daniel Waldram,
{\em Non-standard embedding and five-branes in heterotic M-Theory},
arXiv:hep-th/9808101, Phys.Rev. D59 (1999) 106005\\
Ron Donagi, Andre Lukas, Burt A. Ovrut, Daniel Waldram,
{\em Non-Perturbative Vacua and Particle Physics in M-Theory},
arXiv:hep-th/9811168, JHEP 9905 (1999) 018\\ 
Ron Donagi, Andre Lukas, Burt A. Ovrut, Daniel Waldram,
{\em Holomorphic Vector Bundles and Non-Perturbative Vacua in M-Theory},
arXiv:hep-th/9901009, JHEP 9906:034,1999\\ 
Andre Lukas, Burt A. Ovrut, Daniel Waldram,
{\em Heterotic M-Theory Vacua with Five-Branes},
arXiv:hep-th/9903144, Fortsch.Phys.48:167-170,2000\\
Burt A. Ovrut,
{\em N=1 Supersymmetric Vacua in Heterotic M-Theory},
arXiv:hep-th/9905115,
Lectures presented at the APCTP Third Winter School on "Duality in Fields and Strings", 
February, 1999, Cheju Island, Korea\\ 
Andre Lukas, Burt A. Ovrut,
{\em Symmetric Vacua in Heterotic M-Theory},
arXiv:hep-th/9908100\\
Ron Donagi, Burt A. Ovrut, Tony Pantev, Daniel Waldram,
{\em Standard Models from Heterotic M-theory},
arXiv:hep-th/9912208, Adv.Theor.Math.Phys. 5 (2002) 93-137\\ 
Ron Donagi, Burt A. Ovrut, Tony Pantev, Daniel Waldram,
{\em Standard Model Vacua in Heterotic M-Theory},
arXiv:hep-th/0001101,
Talk given at STRINGS'99, Potsdam, Germany, July 19-24, 1999 \\
Ron Donagi, Burt Ovrut, Tony Pantev, Dan Waldram,
{\em Standard-Model Bundles on Non-Simply Connected Calabi--Yau Threefolds},
arXiv:hep-th/0008008, JHEP 0108:053,2001\\ 
Burt A. Ovrut, 
{\em Lectures on Heterotic M-Theory},
arXiv:hep-th/0201032,
Lectures presented at the TASI 2000 
School on Strings, Branes and Extra Dimensions, Boulder, Co, June 3-29, 2001\\ 
Burt A. Ovrut, Tony Pantev, Rene Reinbacher,
{\em Torus-Fibered Calabi-Yau Threefolds with Non-Trivial Fundamental Group},
arXiv:hep-th/0212221, JHEP 0305 (2003) 040\\
Burt A. Ovrut, Tony Pantev, Rene Reinbacher,
{\em Invariant Homology on Standard Model Manifolds},
arXiv:hep-th/0303020, JHEP 0401 (2004) 059\\ 
Ron Donagi, Burt A.Ovrut, Tony Pantev, Rene Reinbacher,
{\em SU(4) Instantons on Calabi-Yau Threefolds with $Z_2$ x $Z_2$ Fundamental Group},
arXiv:hep-th/0307273, JHEP0401:022,2004\\ 
Ron Donagi, Yang-Hui He, Burt A. Ovrut, Rene Reinbacher,
{\em Moduli Dependent Spectra of Heterotic Compactifications},
arXiv:hep-th/0403291,Phys.Lett.B598:279-284,2004\\ 
Ron Donagi, Yang-Hui He, Burt A. Ovrut, Rene Reinbacher,
{\em The Particle Spectrum of Heterotic Compactifications},
arXiv:hep-th/0405014,JHEP0412:054,2004\\ 
Ron Donagi, Yang-Hui He, Burt Ovrut, Rene Reinbacher,
{\em Higgs Doublets, Split Multiplets and Heterotic $SU(3)_C x SU(2)_L x U(1)_Y$ Spectra},
arXiv:hep-th/0409291,Phys.Lett. B618 (2005) 259-264\\ 
Volker Braun, Burt A. Ovrut, Tony Pantev, Rene Reinbacher,
{\em Elliptic Calabi-Yau Threefolds with $Z_3 x Z_3$ Wilson Lines},
arXiv:hep-th/0410055, JHEP0412:062,2004\\ 
Ron Donagi, Yang-Hui He, Burt A. Ovrut, Rene Reinbacher,
{\em The Spectra of Heterotic Standard Model Vacua},
arXiv:hep-th/0411156, JHEP0506:070,2005\\
Volker Braun, Yang-Hui He, Burt A. Ovrut, Tony Pantev,
{\em A Heterotic Standard Model},
arXiv:hep-th/0501070, Phys.Lett.B618:252-258,2005\\ 
Volker Braun, Yang-Hui He, Burt A. Ovrut, Tony Pantev,
{\em A Standard Model from the E8 x E8 Heterotic Superstring},
arXiv:hep-th/0502155, JHEP 0506:039,2005\\
Volker Braun, Yang-Hui He, Burt A. Ovrut, Tony Pantev,
{\em Vector Bundle Extensions, Sheaf Cohomology, and the Heterotic Standard Model},
arXiv:hep-th/0505041, Adv.Theor.Math.Phys.10:4,2006\\ 
Volker Braun, Yang-Hui He, Burt A. Ovrut, Tony Pantev,
{\em The Exact MSSM Spectrum from String Theory},
arXiv:hep-th/0512177, JHEP 0605:043,2006\\ 
Volker Braun, Yang-Hui He, Burt A. Ovrut,
{\em Yukawa Couplings in Heterotic Standard Models},
arXiv:hep-th/0601204, JHEP 0604:019,2006\\ 
Volker Braun, Yang-Hui He, Burt A. Ovrut,
{\em Stability of the Minimal Heterotic Standard Model Bundle},
arXiv:hep-th/0602073, JHEP0606:032,2006\\ 
Michael Ambroso, Volker Braun, Burt A. Ovrut,
{\em Two Higgs Pair Heterotic Vacua and Flavor-Changing Neutral Currents},
arXiv:0807.3319, JHEP0810:046,2008\\
Lara B. Anderson, James Gray, Andre Lukas, Burt Ovrut,
{\em The Edge Of Supersymmetry: Stability Walls in Heterotic Theory},
arXiv:0903.5088, Phys.Lett.B677:190-194,2009\\ 
Lara B. Anderson, James Gray, Andre Lukas, Burt Ovrut,
{\em Stability Walls in Heterotic Theories},
arXiv:0905.1748, JHEP 0909:026,2009\\ 
Lara B. Anderson, James Gray, Burt Ovrut, 
{\em Yukawa Textures From Heterotic Stability Walls},
arXiv:1001.2317\\
Lara B. Anderson, Volker Braun, Robert L. Karp, Burt A. Ovrut, 
{\em Numerical Hermitian Yang-Mills Connections and Vector Bundle Stability in Heterotic Theories}
arXiv:1004.4399\\
Michael Ambroso, Burt A. Ovrut, 
{\em The Mass Spectra, Hierarchy and Cosmology of B-L MSSM Heterotic Compactifications},
arXiv:1005.5392\\
Ron Y. Donagi, 
{\em Taniguchi Lecture on Principal Bundles on Elliptic Fibrations},
arXiv:hep-th/9802094\\
Vincent Bouchard, Ron Donagi,
{\em An SU(5) Heterotic Standard Model}
arXiv:hep-th/0512149, Phys.Lett.B633:783-791,2006\\ 
Vincent Bouchard, Mirjam Cvetic, Ron Donagi,
{\em Tri-linear Couplings in an Heterotic Minimal Supersymmetric Standard Model},
arXiv:hep-th/0602096, Nucl.Phys.B745:62-83,2006\\ 
Ron Donagi, Rene Reinbacher, Shing-Tung-Yau, 
{\em Yukawa Couplings on Quintic Threefolds},
arXiv:hep-th/0605203\\
Vincent Bouchard, Ron Donagi,
{\em On a class of non-simply connected Calabi-Yau threefolds},
arXiv:0704.3096, Comm. Numb. Theor. Phys. 2 (2008) 1-61\\
Vincent Bouchard, Ron Donagi,
{\em On heterotic model constraints},
arXiv:0804.2096, JHEP 0808:060,2008\\ 
Anthony Bak, Vincent Bouchard, Ron Donagi,
{\em Exploring a new peak in the heterotic landscape},
arXiv:0811.1242, JHEP 06 (2010) 108, pp.1-31\\
Yang-Hui He, Maximilian Kreuzer, Seung-Joo Lee, Andre Lukas,
{\em Heterotic Bundles on Calabi-Yau Manifolds with Small Picard Number}
arXiv:1108.1031\\
Yang-Hui He,
{\em An Algorithmic Approach to Heterotic String Phenomenology},
arXiv:1001.2419, Mod. Phys. Lett. A, Vol. 25, No. 2 (2010) pp. 79-90\\ 
Lara B. Anderson, James Gray, Yang-Hui He, Andre Lukas,
{\em Exploring Positive Monad Bundles And A New Heterotic Standard Model},
arXiv:0911.1569\\ 
Yang-Hui He, Seung-Joo Lee, Andre Lukas,
{\em Heterotic Models from Vector Bundles on Toric Calabi-Yau Manifolds},
arXiv:0911.0865, JHEP 1005:071,2010\\ 
Lara B. Anderson, James Gray, Dan Grayson, Yang-Hui He, Andre Lukas,
{\em Yukawa Couplings in Heterotic Compactification},
arXiv:0904.2186, Commun.Math.Phys.297:95-127,2010\\ 
Maxime Gabella, Yang-Hui He, Andre Lukas,
{\em An Abundance of Heterotic Vacua},
arXiv:0808.2142, JHEP0812:027,2008\\
Lara B. Anderson, Yang-Hui He, Andre Lukas,
{\em Monad Bundles in Heterotic String Compactifications},
arXiv:0805.2875, JHEP 0807:104,2008\\
Lara B. Anderson, Yang-Hui He, Andre Lukas,
{\em Heterotic Compactification, An Algorithmic Approach},
arXiv:hep-th/0702210, JHEP 0707:049,2007\\ 
Gottfried Curio,
{\em Higgs Multiplets in Heterotic GUT Models},
arXiv:1108.5610\\
Bjorn Andreas, Gottfried Curio, 
{\em On the Existence of Stable bundles with prescribed Chern classes on Calabi-Yau threefolds},
arXiv:1104.3435\\
Bjorn Andreas, Gottfried Curio, 
{\em Spectral Bundles and the DRY-Conjecture},
arXiv:1012.3858\\
Bjorn Andreas, Gottfried Curio,
{\em On possible Chern Classes of stable Bundles on Calabi-Yau threefolds},
arXiv:1010.1644, J.Geom.Phys.61:1378-1384,2011\\ 
Bjorn Andreas, Gottfried Curio,
{\em Deformations of Bundles and the Standard Model},
arXiv:0706.1158, Phys.Lett.B655:290-293,2007\\ 
Bjorn Andreas, Gottfried Curio,
{\em Extension Bundles and the Standard Model},
arXiv:hep-th/0703210, JHEP0707:053,2007\\ 
Bjorn Andreas, Gottfried Curio,
{\em Heterotic Models without Fivebranes}
arXiv:hep-th/0611309, J.Geom.Phys.57:2136-2145,2007\\
Bjorn Andreas, Gottfried Curio, 
{\em Invariant Bundles on $B$-fibered Calabi-Yau Spaces and the Standard Model},
arXiv:hep-th/0602247\\
Gottfried Curio,
{\em Standard Model bundles of the heterotic string},
arXiv:hep-th/0412182 Int.J.Mod.Phys. A21 (2006) 1261-1282\\
Bjorn Andreas, Gottfried Curio, Albrecht Klemm,
{\em Towards the Standard Model spectrum from elliptic Calabi-Yau},
arXiv:hep-th/9903052, Int.J.Mod.Phys. A19 (2004) 1987\\
G. Curio, 
{\em Chiral matter and transitions in heterotic string models},
arXiv:hep-th/9803224, Phys.Lett. B435 (1998) 39-48.

\item
\label{het mod}
Ron Donagi, Burt A. Ovrut, Daniel Waldram,
{\em Moduli Spaces of Fivebranes on Elliptic Calabi-Yau Threefolds},
arXiv:hep-th/9904054, JHEP 9911 (1999) 030\\ 
Burt A. Ovrut, Tony Pantev, Jaemo Park,
{\em Small Instanton Transitions in Heterotic M-Theory},
arXiv:hep-th/0001133, JHEP 0005 (2000) 045\\ 
Eduardo Lima, Burt Ovrut, Jaemo Park, René Reinbacher,
{\em Non-Perturbative Superpotentials from Membrane Instantons in Heterotic M-Theory},
arXiv:hep-th/0101049, Nucl.Phys. B614 (2001) 117-170\\ 
Eduardo Lima, Burt Ovrut, Jaemo Park,
{\em Five-Brane Superpotentials in Heterotic M-Theory},
arXiv:hep-th/0102046, Nucl.Phys. B626 (2002) 113-164\\ 
Evgeny Buchbinder, Ron Donagi, Burt A. Ovrut,
{\em Vector Bundle Moduli and Small Instanton Transitions},
arXiv:hep-th/0202084, JHEP 0206 (2002) 054\\ 
Evgeny I. Buchbinder, Ron Donagi, Burt A. Ovrut,
{\em Superpotentials for Vector Bundle Moduli},
arXiv:hep-th/0205190, Nucl.Phys.B653:400-420,2003\\
Evgeny I. Buchbinder, Ron Donagi, Burt A. Ovrut,
{\em Vector Bundle Moduli Superpotentials in Heterotic Superstrings and M-Theory},
arXiv:hep-th/0206203, JHEP 0207 (2002) 066\\
Yang-Hui He, Burt A. Ovrut, Rene Reinbacher,
{\em The Moduli of Reducible Vector Bundles},
arXiv:hep-th/0306121,JHEP 0403 (2004) 043\\ 
Evgeny I. Buchbinder, Burt A. Ovrut, Rene Reinbacher,
{\em Instanton Moduli in String Theory},
arXiv:hep-th/0410200, JHEP0504:008,2005\\ 
Volker Braun, Yang-Hui He, Burt A. Ovrut, Tony Pantev,
{\em Heterotic Standard Model Moduli},
arXiv:hep-th/0509051, JHEP 0601:025,2006\\ 
Volker Braun, Yang-Hui He, Burt A. Ovrut, Tony Pantev,
{\em Moduli Dependent mu-Terms in a Heterotic Standard Model},
arXiv:hep-th/0510142, JHEP 0603:006,2006\\ 
Volker Braun, Burt A. Ovrut,
{\em Stabilizing Moduli with a Positive Cosmological Constant in Heterotic M-Theory},
arXiv:hep-th/0603088, JHEP0607:035,2006\\ 
Volker Braun, Evgeny I. Buchbinder, Burt A.Ovrut,
{\em Dynamical SUSY Breaking in Heterotic M-Theory},
arXiv:hep-th/0606166, Phys.Lett.B639:566-570,2006\\ 
Volker Braun, Evgeny I. Buchbinder, Burt A. Ovrut,
{\em Towards Realizing Dynamical SUSY Breaking in Heterotic Model Building},
arXiv:hep-th/0606241, JHEP 0610:041,2006\\ 
James Gray, Andre Lukas, Burt Ovrut,
{\em Perturbative Anti-Brane Potentials in Heterotic M-theory},
arXiv:hep-th/0701025, Phys.Rev.D76:066007,2007\\ 
Volker Braun, Maximilian Kreuzer, Burt A. Ovrut, Emanuel Scheidegger,
{\em Worldsheet Instantons, Torsion Curves, and Non-Perturbative Superpotentials},
arXiv:hep-th/0703134, Phys.Lett.B649:334-341,2007\\
Volker Braun, Maximilian Kreuzer, Burt A. Ovrut, Emanuel Scheidegger,
{\em Worldsheet Instantons and Torsion Curves, Part A: Direct Computation},
arXiv:hep-th/0703182, JHEP0710:022,2007\\
Volker Braun, Maximilian Kreuzer, Burt A. Ovrut, Emanuel Scheidegger,
{\em Worldsheet Instantons and Torsion Curves, Part B: Mirror Symmetry},
arXiv:0704.0449, JHEP0710:023,2007\\ 
James Gray, André Lukas, Burt Ovrut,
{\em Flux, Gaugino Condensation and Anti-Branes in Heterotic M-theory},
arXiv:0709.2914, Phys.Rev.D76:126012,2007\\
Michael Ambroso, Burt Ovrut,
{\em The B-L/Electroweak Hierarchy in Heterotic String and M-Theory},
arXiv:0904.4509, JHEP 0910:011, 2009\\ 
Michael Ambroso, Burt Ovrut, 
{\em The B-L/Electroweak Hierarchy in Smooth Heterotic Compactifications},
arXiv:0910.1129\\
T. Brelidze, B. Ovrut, 
{\em B-L Cosmic Strings in Heterotic Standard Models}, 
arXiv:1003.0234\\
Lara B. Anderson, James Gray, Andre Lukas, Burt Ovrut,
{\em Stabilizing the Complex Structure in Heterotic Calabi-Yau Vacua},
arXiv:1010.0255, JHEP 1102:088,2011\\ 
Lara B. Anderson, James Gray, Burt Ovrut, 
{\em Transitions in the Web of Heterotic Vacua},
arXiv:1012.3179\\
Lara B. Anderson, James Gray, Andre Lukas, Burt Ovrut,
{\em Stabilizing All Geometric Moduli in Heterotic Calabi-Yau Vacua},
arXiv:1102.0011, Phys.Rev.D83:106011,2011\\
Lara B. Anderson, James Gray, Andre Lukas, Burt Ovrut, 
{\em The Atiyah Class and Complex Structure Stabilization in Heterotic Calabi-Yau Compactifications},
arXiv:1107.5076\\
Gottfried Curio, Ron Y. Donagi,
{\em Moduli in N=1 heterotic/F-theory duality},
arXiv:hep-th/9801057, Nucl.Phys. B518 (1998) 603-631\\ 
Gottfried Curio, 
{\em On the Heterotic World-sheet Instanton Superpotential and its individual Contributions},
arXiv:1006.5568\\
Gottfried Curio,
{\em Perspectives on Pfaffians of Heterotic World-sheet Instantons},
arXiv:0904.2738, JHEP 0909:131,2009\\ 
Gottfried Curio,
{\em World-sheet Instanton Superpotentials in Heterotic String theory and their Moduli Dependence},
arXiv:0810.3087, JHEP 0909:125,2009\\

\end{enumerate}
\end{document}